\documentclass[]{aa}

\usepackage{graphicx}
\usepackage{txfonts}
\usepackage[colorlinks, linkcolor={red}, citecolor={blue}, urlcolor={magenta}]{hyperref}
\begin{document}

\titlerunning{Spectroscopy and polarimetry of SDDS J1004+4112}
\authorrunning{L. \v C. Popovi\'c et al.}

   \title{Spectroscopy and polarimetry of  the gravitationally  lensed quasar SDSS J1004+4112 with the 6m SAO RAS telescope}

   \author{L. \v C. Popovi\'c
   \inst{1} 
   \and
   V. L. Afanasiev
   \inst{2}
   \and 
   A. Moiseev
   \inst{2,3} 
   \and
   A. Smirnova
   \inst{2} 
   \and
   S. Simi\'c
   \inst{4} 
   \and \\
   Dj. Savi\'c
   \inst{1,5} 
   \and
   E. G. Mediavilla
   \inst{6,7} 
   \and
   C. Fian
   \inst{6}
   }

    \institute{Astronomical Observatory, Volgina 7, 11000 Belgrade, Serbia \\
    \email{lpopovic@aob.rs}
    \and
    Special Astrophysical Observatory of the Russian AS, Nizhnij Arkhyz, Karachaevo-Cherkesia 369167, Russia \\ 
    \and
    Space Research Institute, Russian Academy of Sciences, Profsoyuznaya ul. 84/32, Moscow 117997, Russia\\
    \and 
    Faculty of Sciences, University of Kragujevac, Radoja Domanovica 12, 34000 Kragujevac, Serbia\\
    \and 
    Universit\'e de Strasbourg, CNRS, Observatoire Astronomique de Strasbourg, UMR 7550, 11 rue de lUniversit, F-67000 Strasbourg,
    France\\ 
    \and  
    Instituto de Astrof\'isica de Canarias, V\'a L\'actea S/N, La Laguna 38200, Tenerife, Spain\\
    \and 
    Departamento de Astrof\'isica, Universidad de la Laguna, La Laguna 38200, Tenerife, Spain
    }
   
   \date{Received September 15, 1996; accepted March 16, 1997}


  \abstract
   {We present new spectroscopic and polarimetric observations of  the gravitational lens
SDSS J1004+4112 taken with the 6m telescope of the Special Astrophysical Observatory (SAO, Russia).}
   {In order to explain the variability that is observed only in the blue wing of the C IV emission line, corresponding to image A, we analyze the spectroscopy and polarimetry of the four images of the lensed system.}
   {Spectra of the four images were taken in 2007, 2008, and 2018, and polarization was measured in the period 2014-2017. Additionally, we modeled the microlensing effect in the polarized light, assuming that the source of polarization is the equatorial scattering in the inner part of the torus.}
   {We find that a blue enhancement in the CIV line wings affects component A   in all three epochs.  We  also find that the UV continuum of component D was amplified in the period 2007-2008, and that the red wings  of CIII] and CIV appear brighter in D than in the other three components. We report significant changes in the polarization parameters of image D, which can be explained by microlensing.Our simulations of microlensing of an equatorial scattering region in the dusty torus can qualitatively explain the observed changes in the polarization  degree and angle of image D. We do not detect significant variability in the polarization parameters of the other images (A, B, and C), although the averaged values of  the polarization degree and angle are different for the different images.}
   {Microlensing of a broad line region model including a compact outflowing component can qualitatively explain the CIV blue wing enhancement (and variation) in component A. However, to confirmed this hypothesis, we need additional spectroscopic observation in future.}


   \maketitle
%

\section{Introduction}

A lensed quasar is an active galactic nucleus (AGN) that is assumed to have a  central supermassive  black hole, surrounded by an accretion disk (which emits mostly in the X-ray from the inner part, but also in the UV/optical continuum in the outer disk part). The X/UV emitted radiation from the accretion disk  photoionizes the surrounding matter, generating  a region that is able to emit (in the process of recombination) broad emission lines,  the so-called broad line region (BLR). Farther out, a torus-like dust region is thought to lie that mostly emits in the infrared \citep[see, e.g.,][]{st12a,net15}.

The different  regions of the AGN have emission peaks at different wavelengths, and they have different dimensions; microlensing magnification therefore affects them differently \citep[see, e.g.,][]{jov08}. This  can explain the  variations observed  in the spectrum of a  microlensed image \citep[chromatic microlensing, see, e.g.,][]{pop05}.

Variations in the spectra of a lensed quasar image that are due to microlensing can be used to explore the innermost structure of lensed quasars \citep[see, e.g.,][etc.]{pop01,ab02,pop05,sl07,bl11,fi16,fi18}

Spectroscopy has been used in several papers  to constrain the  sizes of different emission regions in lensed quasars, from $\gamma$-ray \citep[see e.g.][etc.]{to03,do11,ne15,vo16}, X-ray \citep[see e.g.][etc.]{ch02,pop01a,pop03,dai03,dai04,pop06,ot06,ch12,ch13,kr17,ch17}, and UV/optical \citep[see][e.g.]{pop01,ab02,pop05,ab07,mo12,sl12,br17,fi18} to the infrared \citep[see, e.g.,][]{st12,mc13,sl13,vi16}.

One of  the greatest interests is to constrain the innermost structure of AGN \citep[see, e.g.,][]{ji14,br17,hu17} and especially the BLR \citep[][]{pop01,ab02,sl12,gu13,br17,hu17}, because this region is assumed to be  relatively close to the central  supermassive black hole, and consequently, the broad  emission lines can be used to measure the masses of central black holes \citep[see][]{pet14,med18}. However, to use the broad lines as a tool for black hole mass measurements,  the kinematics and dimensions of the BLR have to be knonw. These can be  studied from the impact of microlensing \citep[see, e.g.,][etc]{pop01,ab02,br17,hu17,fi18}.\\

Additionally, spectropolarimetric observations can give  useful information about the quasar structure in general \citep[see, e.g.,][etc.]{sm04,af14,af18}, especially in the case of lensed quasars \citep[see, e.g.,][etc.]{hu98,bl00,hl07,hu15}. The polarization in spectra of lensed quasars can give more information about the scattering region (assumed to be a torus) as well as about the kinematics of the BLR \citep[see, e.g.,][]{hu15}. Unfortunately, the images of lensed quasars are faint sources, and  in  most cases, the images are very close such that they cannot be resolved and observed in spectropolarimetric mode. However, the broad band polarization of the total image is easier to carry out in order to study the changes in the polarization parameters (Stokes Q and U parameters) in a microlensing event.

Here we present new spectroscopic and polarimetric observations of the lensed quasar  SDSS J1004+4112, a four-image system with source redshift $z_s$=1.734 and  lens redshift $z_l$=0.58 \citep[][]{in03,og04}. This system exhibits an unusually large separation between images of even 15.0$^{\prime\prime}$ \citep[for more details, see][]{in03,og04,wi04}. SDSS J1004+4112 is also interesting because of the variability of the broad emission lines in component A. The variability is observed in the blue wing of CIV and is not observed in the continuum or in the low-ionization lines \citep[see, e.g.,][]{ri04,la06,gr06,go06,fi18}.

First, \cite{ri04} reported  a 28-day-long amplification event in the broad emission lines of component A that was observed in 2003. A second enhancement observed in 2004 was reported by \cite{go06}
and  confirmed by  \cite{la06}. These events are difficult to explain in terms of gravitational microlensing because the expected  follow-up amplification in another part of the line profiles has not been  detected \citep[as expected in the microlensing of a disk-line profile, e.g.., see][]{pop01}. Additionally, there is  no detection of a continuum amplification that should be present during  the  BLR microlensing event because the BLR surrounds the compact continuum source. An alternative explanation of the  J1004+4112 line variability  was  given by \citet[][]{gr06}, who  assumed that the variability in the C IV blue wing of  component A is caused by the absorption rate that is coming  from matter surrounding the QSO center.  In this case, the difference between A and  the other images  could be due to the small viewing angle  differences that  result in slightly different light paths through the intervening matter. However, this explanation has been ruled out because the model predicts a significant X-ray absorption in components B, C, and D, which has not been observed \citep[see][]{la06}.

The question of the  origin of the CIV A component variation  in J1004+4112 remains unsolved, and it motivates  us to continue  observing this lensed quasar.  We obtained spectroscopic  observations of J1004+4112 in three different epochs from 2007 to 2018 and polarimetric observations in the period 2014--2017 using the 6m telescope of the Special Astrophysical Observatory (SAO).

The paper is organized as  follows: In \S 2 we describe our observations, and in \S 3 we explain the methodology. The results are shown  in  \S 4, and the main conclusions are summarized in \S 5.

\begin{table*}
\begin{center}
\caption{Log of spectral observations of the lensed quasar J1004+4112 with the 6m telescope.}
\label{speclog}
\begin{tabular}{@{}lllllll@{}}
\hline
Object    & Device &Date & T$_{exp}$ & Sp. Range & Sp. Res. & Seeing  \\
              &             &  & (sec)     & (\AA)     & (\AA)    & (arcsec)  \\
\hline
J1004+4112 (ABD) & MPFS &2007 May  16 &  7200     &    3800 $\div$ 6100 &   8   & 1.5
\\
J1004+4112 (ABC) & MPFS & 2007 May  17 &  7200     &    3800 $\div$ 6100 &   8   & 1.3
\\
J1004+4112 (AB) & SCORPIO & 2008  Oct  27 &  4800     &    3650 $\div$ 7540   &   10   & 1.1\\
J1004+4112 (CD) & SCORPIO & 2008  Oct  28 &  4800     &    3650 $\div$ 7540   &   10   & 0.9\\
J1004+4112 (AB) & SCORPIO-2 & 2018  Feb  07 &  600     &    3650 $\div$ 7250   &   5   & 1.7\\
J1004+4112 (CD) & SCORPIO-2 & 2018  Feb  07 &  900     &    3650 $\div$ 7250   &   5   & 1.7\\
\hline

\end{tabular}
\end{center}
\end{table*}
\

\

\section{Observations}

\subsection{MPFS observations and  data reduction.}\label{obs}

J1004+4112 was observed with the integral-field MultiPupil Fiber Spectrograph (MPFS)  located at the prime focus of the  6m telescope of the SAO of the Russian Academy of Sciences (SAO RAS). MPFS takes simultaneous spectra from 256 spatial elements (constructed in the shape of square lenses) that form an array of $16\times16$ elements  on the sky with an angular size of 1 arcsec/element \citep[see][]{Afanasiev2001}. Behind each lens an optical fiber directs the light to the spectrograph slit. The sky background spectrum was simultaneously taken with another fiber bundle  placed  at  a distance of $\sim4$ arcmin from the lens array.  The  detector  we used was  an EEV42-40 ($2K\times2K$ pixels)  CCD. Additional information about the observations  is given in Table \ref{speclog}.

\begin{figure}[h]
\centering
\includegraphics[width=8cm]{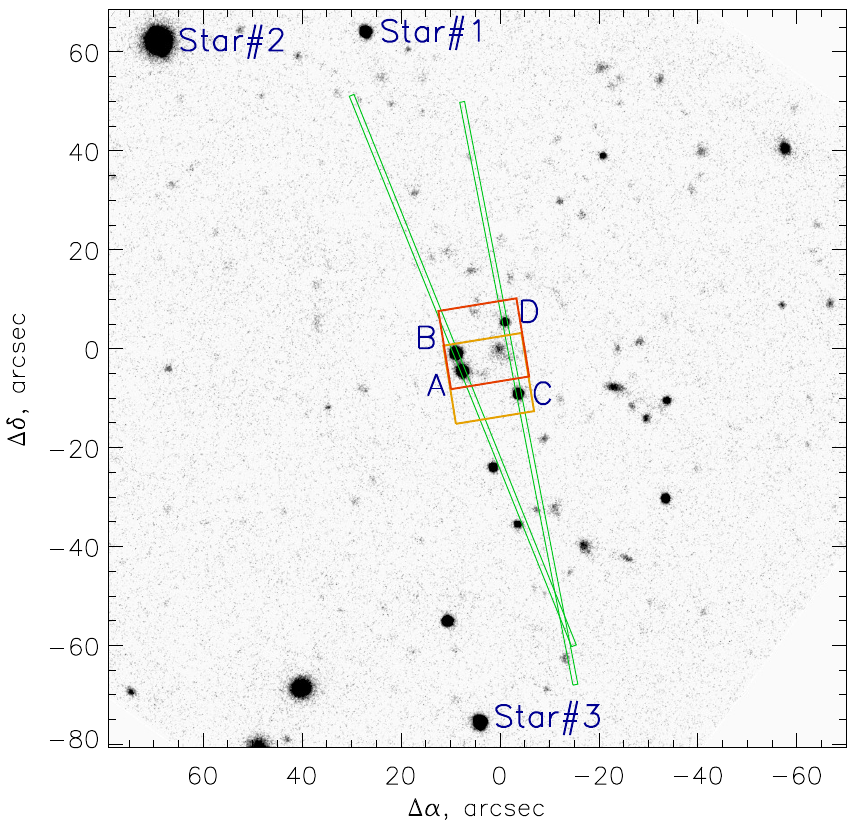}
\caption{Image from the 6m telescope SCORPIO-2   (V-band, seeing $\approx1''$ ) of the field around J1004+4112: four components  of the gravitational lens and three reference stars are labeled. Green lines  represent the position of the spectrograph slits. Red and orange rectangles correspond to the two MPFS fields used during the observations.}
 \label{fig-slit}
\end{figure}

The  data reduction procedure has been described in several papers \citep[see, e.g.,][]{Smirnova2007}. Reduction yields a data cube with an individual spectrum in  each pixel in the $16\times16$ arcsec field. Spectra from spectrophotometric standard stars were used to convert counts into absolute fluxes  ($F_{\lambda}$). We observed the object twice because the angular  distance between images C and D exceeds the MPFS field of view. During the first night, the MPFS array was centered  to collect spectra of images  A, B, and D, and during the second night, we observed images A, B, and C (see  pointings of MPFS in Fig. \ref{fig-slit}). After the primary data reduction, these two cubes were matched and combined into a mosaic cube with a resulting field of view of $18\times24$ arcsec$^2$.

To reproduce spectra of each component, we collected the integrated spectra taking several bright pixels at the position of the image in apertures og 1.5-2 arcsec in radius. The  Hubble Space Telescope (HST) image of this lensed system suggests a nearby object within 2 arcsec of component A \citep[see Fig. 1 in][]{sh05}. In our MPFS observations we were unable to correctly deblend this object.  However, the possible contribution of this galaxy in the integrated C IV emission line is lower than 5\%, according the flux estimation in the MPFS data cube in the corresponding location.

\subsection{Long-slit observations}

The long-slit spectra of the four components of J1004+4112 were observed at the prime focus of the SAO RAS 6m telescope in  October 2008 and  February 2018 with the multi-mode focal reducer SCORPIO \citep[][]{am05} and its improved version SCORPIO-2 \citep{AM11}. A long-slit  with a width of $1$ arcsec was placed along A-B  at   position angle $PA=22^\circ$ and along C-D  at $PA=11^\circ$; see Fig. \ref{fig-slit}.  Both devices provide the same scale  of $0.36$ arcsec per pixel with a similar spectral range  (see Table~\ref{speclog}). The spectral dispersion and resolution were twice better in the 2018  observations than in 2008 because  gratings and detectors were of better quality.

In 2008 we used the  CCD detector   EEV42-40 ($2K\times2K$ pixels), while in 2018 we used an E2V 42-90  detector with a larger number of pixels  ($4.6K\times2K$ pixels). The bias subtraction, geometrical
corrections, flat fielding, wavelength  scale calibration, sky subtraction, and calibration to flux units were performed using the IDL-based software that is briefly described in \citet[][]{am05}.

\begin{table}[h]
\begin{center}
\caption{Log of  polarization  observations of the lensed quasar J1004+4112 with the 6m telescope of SAO RAS.}
\label{pollog}
\begin{tabular}{cccccc}
\hline
Date  &JD-2450000               &T$_{exp}$ & Number     & Seeing  \\
       &  (days)             &      (sec)  & of cycles & (arcsec) \\
\hline
2014 Nov 23& 6984 &   5400  & 10    &  1.8  \\
2014 Nov 28& 6989 &   2880  & 8     &  1.0  \\
2015 Dec 10& 7366 &   5400  & 6     &  2.0  \\
2016 Mar 08& 7455 &   3600  & 6     &  1.4  \\
2016 Apr 05& 7483 &   3600  & 6     &  1.8 \\
2016 Nov 24& 7716 &   3600  & 6     &  1.2  \\
2016 Dec 21& 7743 &   4500  & 5     &  1.5  \\
2017 Jan 22& 7775 &   4500  & 5     &  1.2  \\
\hline
\end{tabular}
\end{center}
\end{table}

The  atmospheric extinction correction for spectrophotometric standards and source was performed in a standard way. The air mass was  taken to be proportional to $\sec(z)$ ($z$ is the zenith length of the object in the time of observations), and  the Rayleigh scattering, as well as atmospheric absorption,  was calculated taking into account the measurements of the spectral atmospheric transparency  at the place where the 6m telescope is located \citep[given by][]{ka78}.

\subsection{Polarization observations}

In the period from 2014 to 2017, we performed  polarimetric observations of J1004+4112 at  eight epochs. The log of observations is given in Table \ref{pollog}.  The dichroic polarizer was used as a polarization analyzer.   The polarization was measured with the method of Fesenkov. This method uses  a series of observations at three fixed  rotation angles of the  analyzer: -60, 0, and + 60 degrees.

The number of cycles,  consisting of three consecutive frames at these angles,  as well as the total exposure  times are shown in Table \ref{pollog}. Images were obtained in the photometric V band of the Johnson system. In Fig. \ref{fig-slit} we mark three reference stars that we used for photometric calibration. 


To find the  zero-point of the polarization angle (PA), we observed the polarization standard HD25443 (with p=5.13\% and  PA=134.2 degrees). The method of image-polarimetry, which accounts for the instrumental polarization, as well as  atmospheric variability is describe in \citet{ai18}, and here we do not repeat it in detail. We only briefly mention that we used a field of view $6^\prime\times  6^\prime$, covering around 50 stars, made histograms of the Stokes parameters Q and U, and found their averaged values as   $<Q>=-0.15\pm0.21$ and $<U>=0.05\pm0.17$. The averaged Stokes parameters are the vector sum of the  interstellar polarization in the direction of the Galactic longitude of  b=27.3 degrees. To extract polarization of a lens component, we subtracted these averaged Stokes parameters from the observed U and Q of the lens component. The interstellar polarization in this direction is  lower than p=0.07\% \citep[see][]{he00}. This method yields results that show that the instrumental polarization is lower than p=0.2\% \citep[see][]{ai18}, which is taken into account in our measurements of the Q and U parameters.

Photometry of each lensed image was observed in the circle-like aperture with diameter of 4$^{\prime\prime}$ centered at the image center with a spatial accuracy of about 0.15$^{\prime\prime}$. The flux measurements were been on the local standard stars (denoted as stars 1-3 in Fig. \ref{fig-slit}). The BVR fluxes of these stars were found using observations of NGC2420.

In the case of image polarimetry, which means differential measurements based on the photometric standards, the influence of atmospheric extinction can be neglected. The atmospheric depolarization was taken into account using the method described in \citet[][]{am12}.

After  measuring the intensities in the three angle-positions of the Polaroid - $I (x, y)_{0}$, $I(x,y)_{-60}$, and $I(x,y)_{+60}$, we  find the total intensity $I$ and  normalized Stokes parameters $Q$ and $U$ at each point of the image with coordinates $(x, y)$, using the following relationships:

\begin{equation}
\begin{array}{c}
I(x,y)=\displaystyle{\frac{2}{3}(I(x,y)_{0}+ I(x,y)_{-60}+I(x,y)_{+60})}
\\
\\
Q(x,y)=\displaystyle{\frac{2I(x,y)_{0}-I(x,y)_{-60}- I(x,y)_{+60}}
{I(x,y)_{0}+ I(x,y)_{-60}+I(x,y)_{+60}}}\\
  \\
U(x,y)=\displaystyle{\frac{\sqrt{3}}{{2}}\frac{I(x,y)_{+60}-I(x,y)_{-60}}
{I(x,y)_{0}+ I(x,y)_{-60}+ I(x,y)_{+60}}}~.\\
\end{array}
\end{equation}\
The degree of polarization $P$  and  the polarization angle  $\varphi$ are calculated as
\begin{equation}
\begin{array}{ccc}
P=\sqrt{Q^2+U^2,}& & \varphi=\varphi_{slit}-{1\over2}\arctan{{U}\over{Q}}+\varphi_0~.\\
\end{array}
\end{equation}
Here $\varphi_{slit}$ is the angle of the vertical direction in the image and $\varphi_0$ is the zero-point, which was determined by observations of polarization standards. The instrumental polarization and depolarization of Earth's atmosphere were taken into account using the method described in \citet{ai18}.

\begin{figure}[]
\centering
\includegraphics[width=8 cm]{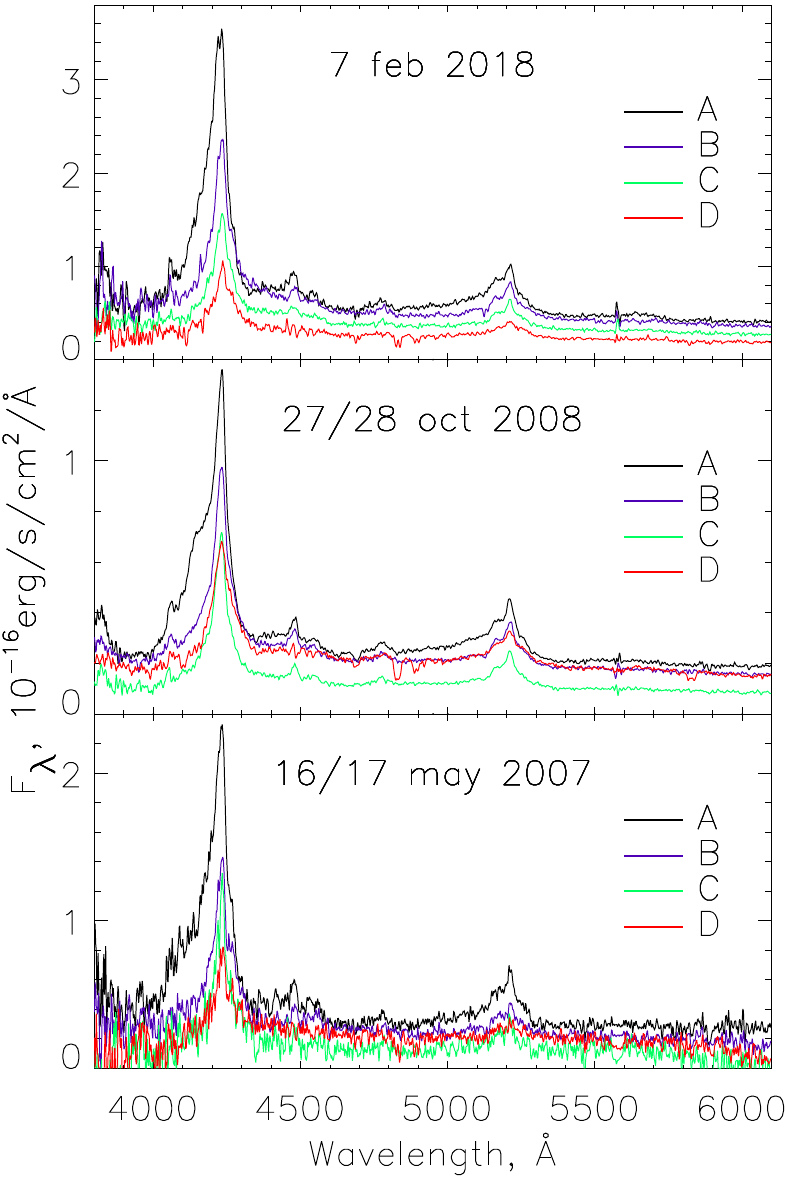}
\caption{Spectra of the four components corresponding to three epochs:  2007 (bottom),  2008 (middle), and  2018 (top).}
\label{fig-l}
\end{figure}

\section{Results}

We took spectra of the four components in three epochs: 2007, 2008, and 2018. Additionally, we observed the polarization of the four images of J1004+4112 twice during 2014, four times in 2016, and one time in 2015 and 2017. We also compared our spectroscopic observations with those published earlier in 2003 \citep[][]{in03}.

\begin{figure*}[t!]
\centering
\includegraphics[width=7cm]{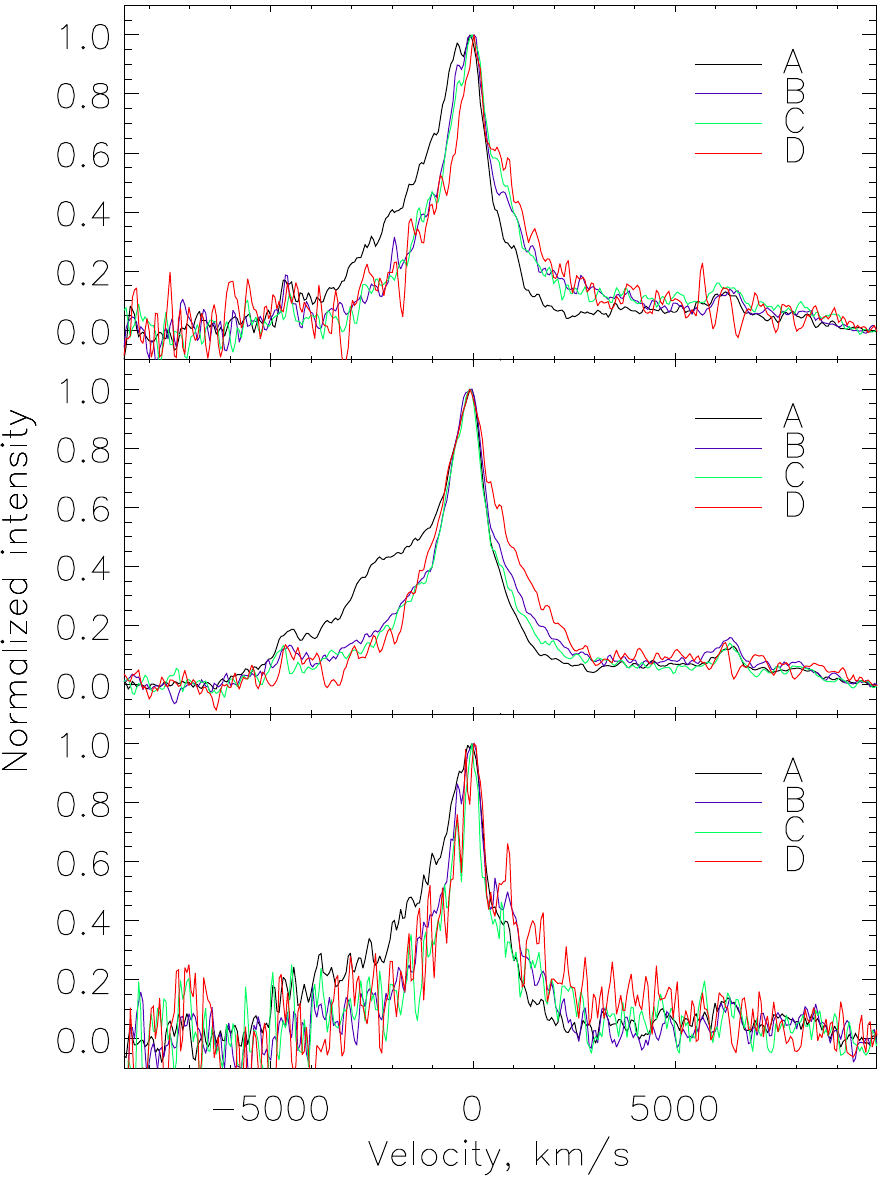}
\includegraphics[width=7cm]{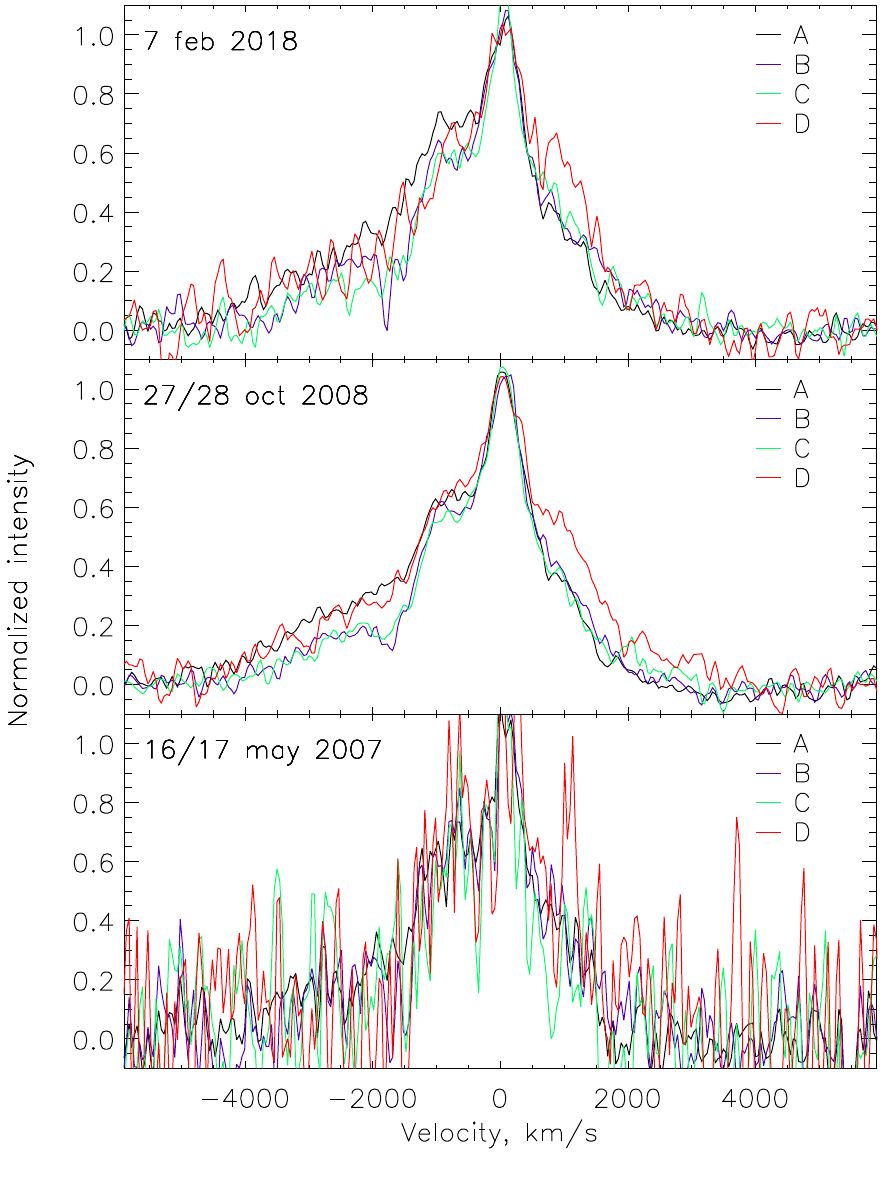}
\caption{ C IV (left) and CIII] (right) emission line profiles of the four images in the three epochs:  2007 (bottom), 
2008 (middle), and  2018 (top).}
\label{fig-l1}
\end{figure*}

\subsection{ Spectroscopic variability}

When we compared the spectra obtained in the three epochs, we find a significant flux increase in the CIV blue line  wing of component A (see Fig. \ref{fig-l}). This enhancement in the CIV blue wing was reported earlier by several authors \citep[see, e.g.,][]{ri04,go06,gr06,mo12,fi18}. The changes in the CIV line profile of the other three components were not significant. We only detect a change in the continuum level of component D, as shown in Fig. \ref{fig-l}. The continuum flux of component D was higher in 2007-2008 than in 2018. At these epochs, the conitnuum level of image D was similar to the continuum level observed in  image B. However, in 2018 the continuum level was significantly lower in component D than in component C. This suggests that a significant microlensing event during 2007 and 2008 may have affected component D (see Fig. \ref{fig-l}). Our error bars in the absolute flux measurements are estimated on the level 10\%-15\%, therefore the observed changes in the D component continuum seem to be the real. 

To explore changes in the line profiles, first we estimated the level of the continuum by fitting a cubic spline function through the windows that covered the following spectral ranges: around 3900\AA,\, 4600\AA,\ and 5400\AA. After this, the continuum was subtracted. The estimated error bars due to  the continuum-subtraction procedure are at 3\%-4\%. The  CIV and C III] lines were normalized to the line peaks and are presented in Fig. \ref{fig-l1}.
 
A comparison of the normalized CIV line profiles of the four components observed in the three epochs (see the left panels in Fig. \ref{fig-l1}) clearly shows a significant flux increase in the CIV blue wing only in component A. It is interesting to note that the red wing of  the C IV line of component A is slightly smaller than in the other three components (see the first two left panels in Fig. \ref{fig-l1}).

\begin{figure}[]
\centering
\includegraphics[width=8cm]{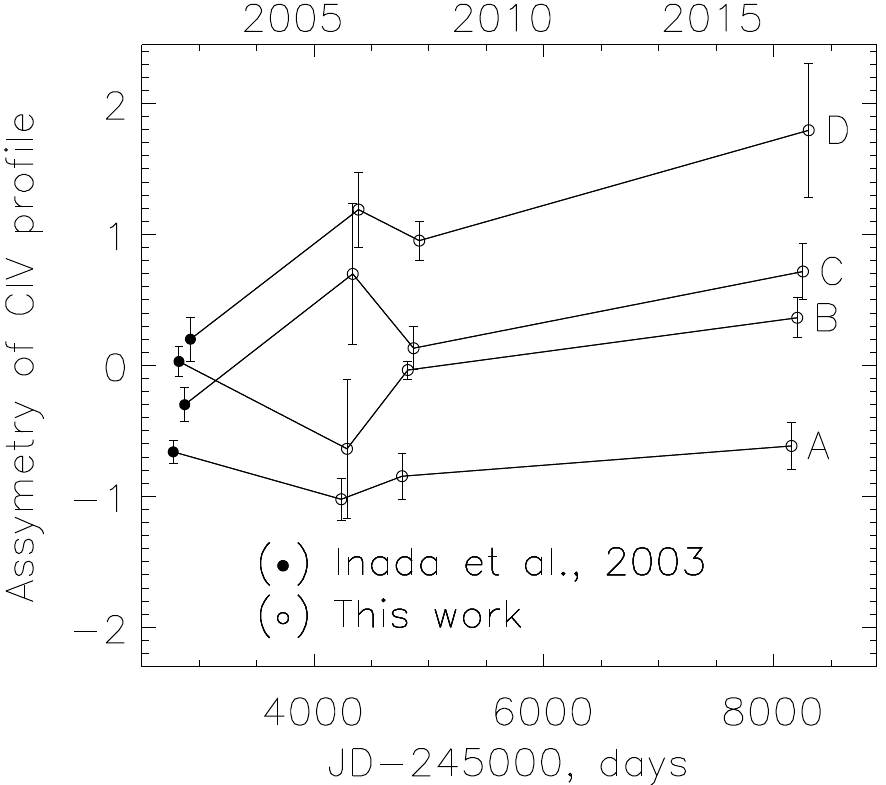}
\caption{Asymmetry coefficients (see text) of the CIV$\lambda$1549 line profile, obtained for each one of the four images at four epochs. 
The data corresponding to 2003 are reproduced from the observations reported by \citet{in03}.}
 \label{fig-as}
 \end{figure}

In the right panels of Fig. \ref{fig-l1} we compare the CIII]$\lambda$1909 line profiles of the four components. They show that the CIII] line profiles of the different images are similar. Only in 2008 does the red wing of component D seem to be  clearly higher (see the middle right panel in Fig. \ref{fig-l1}),  but  an increase may also be present in 2018 (mostly in the lowest velocity fraction of the red wing). This enhancement may be caused by the microlensing event that is detected in the continuum (see Fig. \ref{fig-l}).

\citet[][]{in03} showed the first  spectra with high signal-to-noise ratio (S/N) of SDSS J1004+4112 images that were observed with the Keck telescope. These spectra clearly  show that the C IV line has an asymmetric profile with a different asymmetry in the different images \citep[see][]{in03}. This motivates us to measure the  C IV asymmetry  coefficient ($\gamma$) as \citep[see, e.g.,][]{gm84}
$$\gamma={M_3\over{M_2^{3/2}}},$$
where $M_2$ and $M_3$ are the second and third moments of the line profiles. In Fig.\ref{fig-as} we show the measured  asymmetries of the  C IV lines  for all four images.

To measure the C IV asymmetry, first we explored the influence of  different integration windows around the CIV line center. First we took a window of  $\pm$ 15.5 \AA\ around the line center (which corresponds to a full width at half-maximum, FWHM, that is $\sim 31$ \AA\ or 6000 km s$^{-1}$). This window includes about 80\% of the total line flux, and we found that the asymmetry is $-1.19\pm 0.08$. Then we chose a larger window $\pm$ 25.8 \AA\ around the line center (in total  10000  km s$^{-1}$) that covered 95\%  of the total line flux, and found that in this case, the asymmetry is $-1.01\pm 0.07$. To avoid the influence of an absorption in the blue wing of C IV and a weak emission of  He I (see Fig. \ref{fig-l1}), we chose a wavelength interval  $\sim \pm$ 25.8 \AA\ around the line center to measure the asymmetry.

The second task was to determine the asymmetry and  estimate  the error bars of the C IV measured asymmetry. To do this,  we performed the following procedure: (a) For each C IV line profile we estimated the error bars in the measured asymmetry using the bootstrap method \citep[see][]{ef79}, that is, we first we produced a Monte Carlo random sample of asymmetry estimates of the observed profiles by taking different (random) noises that contribute to the observed line profile errors; (b) After this, we constructed the histogram of the asymmetry where the averaged peak values were taken as the asymmetry and the histogram width as an estimated error of the asymmetry. Additionally, we estimated that the continuum subtraction  contributes to the error bars by around 3-4\%.

Figure \ref{fig-l1} shows that the blue asymmetry of the A component can be clearly detected by comparing the line profiles. The red asymmetry in component D can also be detected, while components B and C seem to have a weaker and insignificant asymmetry. 

We tried to measure the CIII] asymmetry, but we found that the error bars are too high. We therefore cannot give any valid conclusion about CIII] asymmetry.

Figure \ref{fig-as} shows that  the A component  has a blue asymmetry in all epochs.  Additionally, we include the asymmetry coefficients corresponding to the observations in 2003  by \citet{in03} (first point in the plots of Fig. \ref{fig-as}), and we found that in component A, the blue asymmetry of CIV in 2003 was weaker than in our observations. It is interesting to see that the CIV line in the component D has a red asymmetry that is more  prominent in the 2018 observations (see also Fig. \ref{fig-l}).

To summarize, the main results of the spectroscopic observations were the following:
 \begin{itemize}
  \item  There is  an enhancement of  the blue wing of the CIV line of image A in all three epochs, with a maximum flux increase in the blue wing in 2008.
  \item The continuum flux of image D increased in the period of 2007-2008. In 2008, the CIII] red wing of image D also appears to be enhanced.
  \item The blue asymmetry of the CIV profile of the A component is present from 2003 to 2018, while component  D seems to have a red asymmetry.
 \end{itemize}

\subsection{ Polarization variability}

\begin{figure}[t]
\includegraphics[width=8cm]{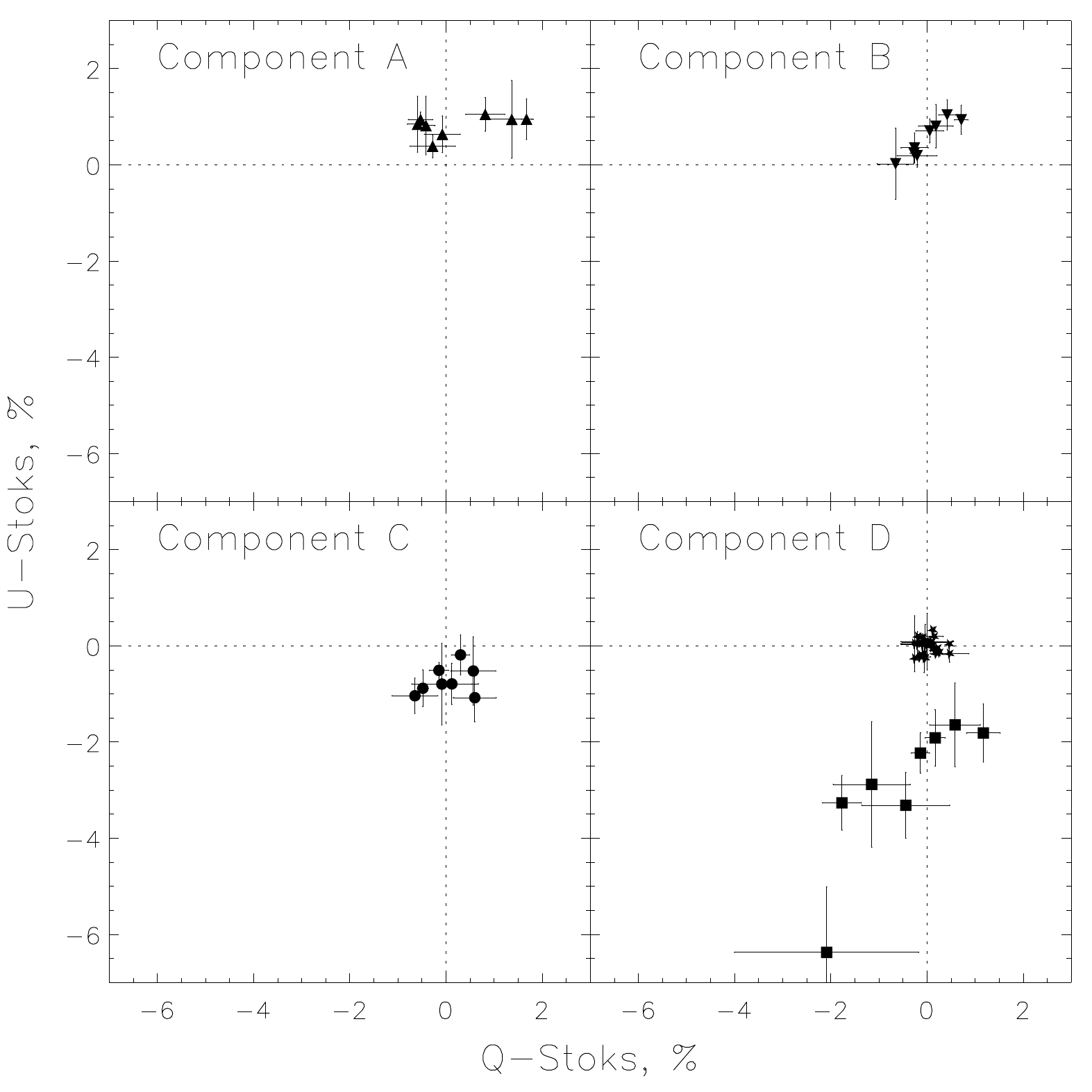}
\caption{ UQ-plane plots for the four components of QSO J1004+4112.  Values of UQ for standard stars with zero polarization are plotted (points concentrated in the center, with zero-zero position) in the panel corresponding to image D, where off-centered points represent the change in UQ parameters of component D.}
\label{fig-pol}
\end{figure}

\begin{figure*}[t]
\centering
\includegraphics[width=16cm]{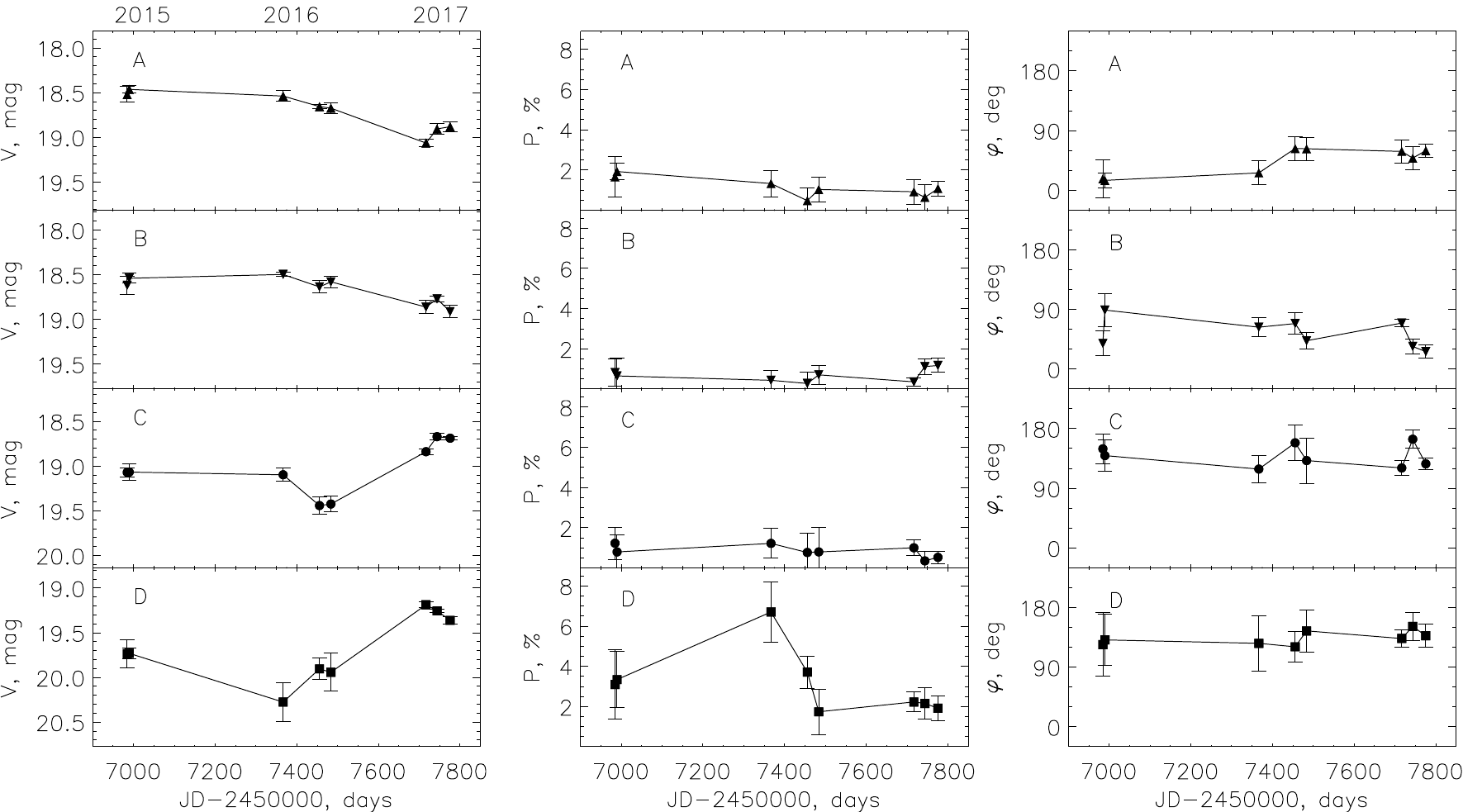}
\caption{Variability in the magnitude (left), polarization (middle), and polarization angle (right) for all four components.}
\label{fig-pol1}
\end{figure*}

The polarized light of J1004+4112 was observed in the period from 2014 to 2017 (see Table \ref{pollog}), covering eight epochs. In Figs. \ref{fig-pol} and \ref{fig-pol1} we present the results. We also give  our measurements of the polarization parameters (for all epochs and the averaged parameters) for the four components in Table \ref{polpar}.

\
In Fig. \ref{fig-pol} we present variations in the U and Q Stokes parameters using the UQ-plane for all images. In the fourth plot (image D) we also show the observed UQ obtained for polarization zero standards (points with zero-zero position on the UQ-plane). The figure shows slight changes in the UQ plane for components A, B, and C, and strong changes can be seen in component D (off-centered points in the UQ plot of component D).

In Table \ref{polpar} we list the observed polarization parameters (Q, U, p, and $\varphi$) for all epochs and their averaged values obtained in the period 2014-2017. The table shows (and also in Fig. \ref{fig-pol}) that component A has an averaged location in the QU plane of  $<Q>=$0.26 and $<U>=$0.82 corresponding to an averaged polarization angle of $\varphi\sim$40 degrees, which is slightly smaller than the polarization angle in component B, and quite different than the angles in components C and D ($<\varphi> \sim$130 degrees).

In Fig. \ref{fig-pol1} we present the variability in the level of polarization (p) and the polarization angle ($\varphi$) during the period 2014-2017: we plot changes in the V magnitude (first panel), the level of polarization in percent (second panel), and the  polarization angle (third panel). Small changes in the magnitude of components A and B are evident, as are changes in the polarization parameters (p and $\varphi$). A strong change in the magnitude and in the polarization parameters is detected in component D. Components A, B, and C show polarization between 0.5\% to 2\%, as expected for type 1 AGNs. The expected level of polarization in type 1 AGNs is $\leq$1\% \citep[see, e.g.,][]{sm04,af18}. While component D shows a higher variability in the  level of polarization in the first three epochs (between 3\% and 7\%). In this period, we also detect flux variability in image D (see the first panel in Fig. \ref{fig-pol1}).


\begin{table}[h]
\begin{center}
\caption{Polarization parameters for the four components of  the gravitational lens J1004+4112 observed in different epochs and their averaged values for the period 2014-2017.}
\label{polpar}
component A\\
\begin{tabular}{|c|c|c|c|c|}
\hline
JD -&$Q$&$U$&$P$& $\varphi$ \\
2450000 &\% & \% & \% & degree \\
\hline

6984&~1.37$\pm$0.63&~0.95$\pm$0.81& 1.66$\pm$1.02&~17$\pm$28\\
6989&~1.67$\pm$0.15&~0.95$\pm$0.42& 1.92$\pm$0.40&~14$\pm$11\\
7366&~0.81$\pm$0.57&~1.05$\pm$0.35& 1.33$\pm$0.65&~26$\pm$18\\
7455&-0.28$\pm$0.67&~0.39$\pm$0.24& 0.48$\pm$0.64&~62$\pm$17\\
7483&-0.59$\pm$0.30&~0.85$\pm$0.58& 1.03$\pm$0.62&~62$\pm$17\\
7716&-0.42$\pm$0.26&~0.82$\pm$0.61& 0.92$\pm$0.61&~58$\pm$17\\
7743&-0.08$\pm$0.52&~0.64$\pm$0.38& 0.64$\pm$0.64&~48$\pm$17\\
7775&-0.53$\pm$0.36&~0.94$\pm$0.16& 1.08$\pm$0.36&~59$\pm$10\\
\hline
averaged&~0.26$\pm$0.78&~0.82$\pm$0.16& 1.13$\pm$0.38&~44$\pm$18\\
\hline
\end{tabular}\\
component B\\
\begin{tabular}{|c|c|c|c|c|}
\hline
JD - &$Q$&$U$&$P$& $\varphi$ \\
2450000      &\% & \% & \% & degree \\
\hline
6984&~0.18$\pm$0.50&~0.81$\pm$0.45& 0.83$\pm$0.68&~38$\pm$18\\
6989&-0.65$\pm$0.54&~0.02$\pm$0.74& 0.66$\pm$0.90&~89$\pm$25\\
7366&-0.26$\pm$0.40&~0.35$\pm$0.31& 0.44$\pm$0.50&~63$\pm$13\\
7455&-0.21$\pm$0.58&~0.19$\pm$0.24& 0.28$\pm$0.58&~68$\pm$16\\
7483&~0.05$\pm$0.41&~0.71$\pm$0.25& 0.71$\pm$0.46&~42$\pm$12\\
7716&-0.27$\pm$0.11&~0.24$\pm$0.19& 0.36$\pm$0.21&~69$\pm$ 5\\
7743&~0.42$\pm$0.25&~1.04$\pm$0.31& 1.12$\pm$0.39&~34$\pm$11\\
7775&~0.71$\pm$0.19&~0.94$\pm$0.30& 1.18$\pm$0.35&~26$\pm$ 9\\
\hline
averaged&~0.01$\pm$0.34&~0.54$\pm$0.34& 0.69$\pm$0.26&~54$\pm$19\\
\hline

\hline
\end{tabular}\\
component C\\
\begin{tabular}{|c|c|c|c|c|}
\hline
JD -&$Q$&$U$&$P$& $\varphi$ \\
 2450000     &\% & \% & \% & degree \\
\hline
6984&~0.60$\pm$0.63& -1.08$\pm$0.50& 1.23$\pm$0.80& 149$\pm$22\\
6989&~0.12$\pm$0.78& -0.79$\pm$0.43& 0.80$\pm$0.85& 139$\pm$23\\
7366&-0.65$\pm$0.67& -1.04$\pm$0.37& 1.22$\pm$0.73& 119$\pm$20\\
7455&~0.57$\pm$0.66& -0.52$\pm$0.71& 0.77$\pm$0.97& 158$\pm$27\\
7483&-0.09$\pm$0.88& -0.79$\pm$0.85& 0.80$\pm$1.22& 131$\pm$34\\
7716&-0.48$\pm$0.16& -0.88$\pm$0.38& 1.00$\pm$0.38& 120$\pm$10\\
7743&~0.30$\pm$0.26& -0.19$\pm$0.41& 0.35$\pm$0.48& 164$\pm$13\\
7775&-0.15$\pm$0.28& -0.51$\pm$0.16& 0.53$\pm$0.31& 126$\pm$ 8\\
\hline
averaged&~0.03$\pm$0.37& -0.72$\pm$0.24& 0.84$\pm$0.25& 138$\pm$14\\
\hline
\end{tabular}\\
component D\\
\begin{tabular}{|c|c|c|c|c|}
\hline
JD - &$Q$&$U$&$P$& $\varphi$ \\
  2450000    &\% & \% & \% & degree \\
 \hline 
6984& -1.15$\pm$1.13& -2.88$\pm$1.30& 3.10$\pm$1.72& 124$\pm$48\\
6989& -0.44$\pm$1.29& -3.31$\pm$0.68& 3.34$\pm$1.39& 131$\pm$39\\
7366& -2.09$\pm$2.71& -6.37$\pm$1.36& 6.70$\pm$1.50& 125$\pm$42\\
7455& -1.77$\pm$0.57& -3.26$\pm$0.57& 3.71$\pm$0.80& 120$\pm$22\\
7483& ~0.58$\pm$0.74& -1.64$\pm$0.87& 1.74$\pm$1.14& 144$\pm$31\\
7716& -0.14$\pm$0.27& -2.23$\pm$0.42& 2.23$\pm$0.49& 133$\pm$13\\
7743& ~1.17$\pm$0.49& -1.81$\pm$0.60& 2.16$\pm$0.78& 151$\pm$21\\
7775& ~0.17$\pm$0.29& -1.91$\pm$0.59& 1.92$\pm$0.62& 137$\pm$17\\ 
\hline
averaged& -0.46$\pm$0.91& -2.9$3\pm$1.04& 3.11$\pm$1.10& 134$\pm$8\\
\hline
\end{tabular}
\end{center}
\end{table}

As a summary of the polarization observations, we can outline the following results:

\begin{itemize}
 \item The averaged polarization parameters for the four images of J1004+4112 are different in each image.
 \item Changes in the polarization parameters of components A, B, and C are not significant,  but component D shows a strong change in the polarization parameters during the 2014-2017 period. Image D shows a higher level of polarization and change in polarization, which is correlated with the change in brightness.
 \item The averaged polarization angles (see Table \ref{polpar}) are different for different components: components A and B seem to have polarization angles of $\sim$40-50 degrees, while components C and D have polarization angles of $\sim$130 degrees.
\end{itemize}

\section{Discussion and interpretation of the observations}

The nature of observed spectroscopic (and polarometric)  variability of an image in a lensed system can be due to intrinsic variability (which is often observed in non-lensed quasars) and microlensing.  Before we discuss and interpret the observed variability in the J1004+4112 lens system, we therefore clarify the nature of the observed variability. 

Time-delay measurements between the  J1004+4112 images show a short time delay between components A and B \citep[B leads A by $\sim$ 40 days, see][]{fo07,fo08}, which is very close to the theoretical predictions \cite[$\sim$ 30 days, see][]{ri04}. The time delay between image C and A is about 820 days (C leads A), and it is longer between image A and D  \citep[D lags A by $>$1250 days, see][]{fo08}. The  similarity between the CIII] profiles for B and C suggests that a change in the line shape of image D (see Fig. \ref{fig-l1}, larger red wing in 2008) caused by intrinsic variability is unlikely because the time delay between B and C is longer than two years. When we compare the contiuum variablilty in Fig. \ref{fig-l}, significant variability is seen in component D (e.g., between May 2007 and October 2008, without any amplification in the C component).  Moreover, \citet{fi18} showed that in the period 2007 -- 2010, component D showed high variability that is caused by gravitational microlensing (see their Fig. 4 and the corresponding discussion in the paper). Especially in the period 2007-2008, the variability of component D was caused by microlensing, as was shown in \citet{fi18}.  

On the other hand, the polarization variability in component D strongly changed by about 4\% -- 5\% (see Fig. \ref{fig-pol1} and Table \ref{polpar}), which is too high to be expected from the intrinsic polarization variability of AGNs \citep[which is around 1\%; see, e.g.,][]{af14,af15,ko16,kok17}. The continuum intrinsic variability in polarization of type 1 AGNs  is probably caused by the change in the accretion disk polarization \citep[see][]{ko16}, and it is not expected to change significantly (below 1-2\%). The maximum contribution of the accretion disk polarization through rediative transfer is about 10\% \citep[][]{ch50}, but as a rule, this is lower in type 1 AGNs at about 0.5-1\%.  A strong change in polarization in component D alone (during the 2014-2017 period) is also unlikely due to intrinsic polarization variability. 

We cannot absolutely rule out a contribution of the intrinsic variability to the observed variability of component D, but it seems that in the observed period where variability in polarization is present, variability caused by microlensing is dominant. Therefore we consider microlensing in this section as the main cause of the observed variability in flux magnification (and polarization) of component D  and of the blue line wing amplification in component A.

\subsection{ Spectroscopic variability}

The detected change in the blue wing of CIV was also reported in  previous papers \citep[see][etc.]{ri04,gr06,la06,mo12,fi18}. To explain the exclusive amplification of the  CIV blue wing in image A (without a similar amplification in the CIII] line of A), we point out two observational facts: (a) The amplification  in the CIV blue wing of component A is not followed by the amplification of the center and/or red wing of the line (as expected in the case of a classical disk-like BLR \citep[see, e.g.,][]{pop01}) or by a magnification of the A image  continuum (neither has it been detected in previous observations \citep[see][]{ri04,gr06,la06,mo12}). We only detect a significant continuum amplification in component D during the first two epochs (see Fig. \ref{fig-l}); (b) In Fig. 3 the C IV red wing of component A appears smaller than the red wings of the other components (especially compared with the red wings of components D and C). This can be a consequence of the normalization to the line maximum if some additional emission contributes to the blue wing and core of the CIV line of image A but not to the red wing (see Fig. \ref{fig-l1}).

\begin{figure}[]
\centering
\includegraphics[width=0.5\textwidth]{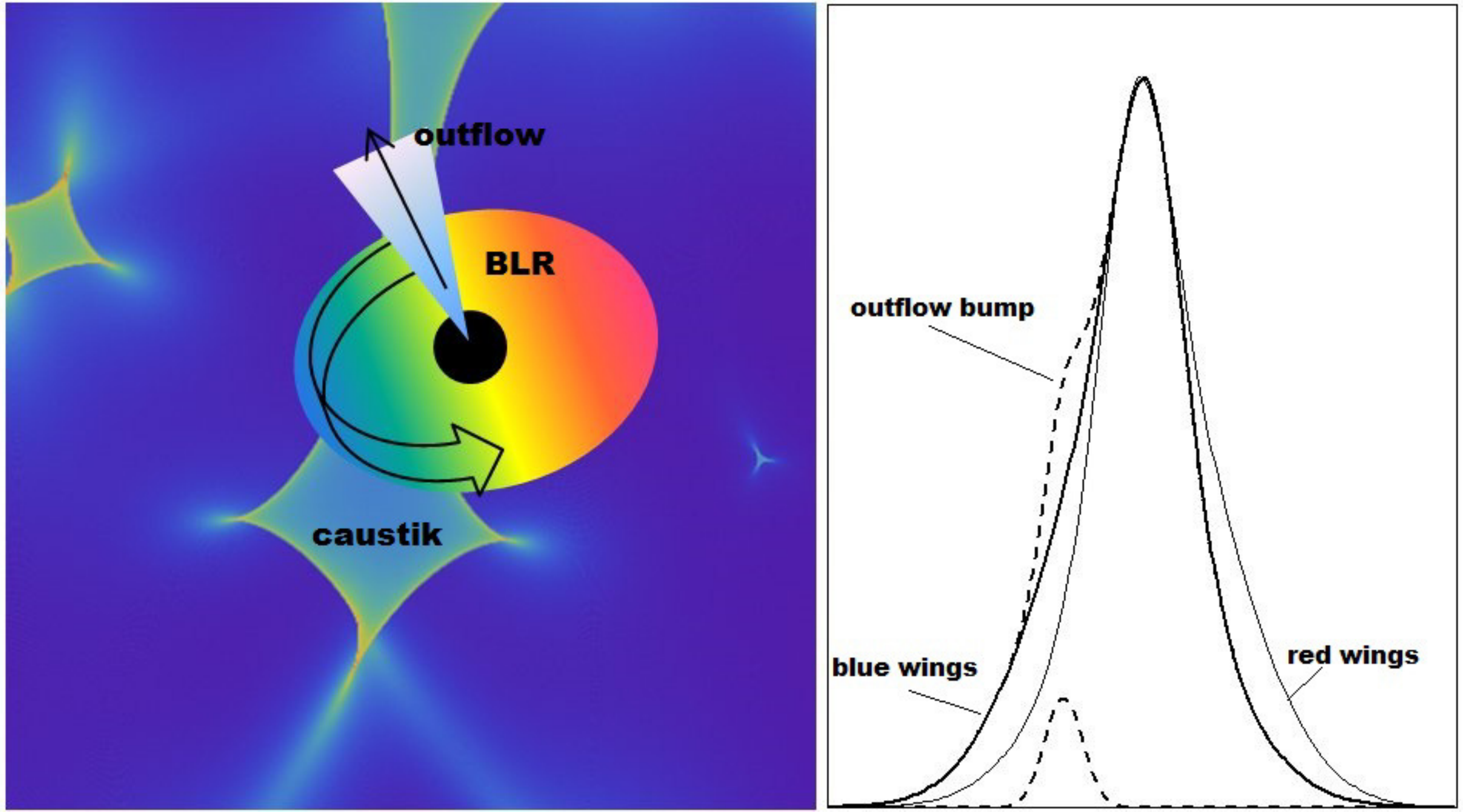}
\caption{Scheme of the caustic crossing of the compact jet-like region (left) that emits a small contribution to the blue wing of the C IV$\lambda$1550  line of QSO J1004+4112 (right). The dimension scale on the left panel is given in arbitrary units, where the disk-like BLR is assumed to be several tens of light days (see text) and the jet-like region is assumed to more compact ($<9$ light days) to be microlensed.} 
\label{fig-mod1}
\end{figure}

To explain these facts, a physical scenario might be considered in which microlensing can magnify only a part of the broad emission lines \citep[see, e.g.,][]{pop01,ab02,ab07}. The C IV BLR dimensions can be estimated using the luminosity of the C IV line and nearby continuum \citep[at $\lambda$1350\AA; see, e.g.,][]{ko06,ka07,tr14,li18,ho19}. To estimate a   continuum at $\lambda$ 1350 \AA\, that is not amplified,  we first calculated the amplification ($A$) for each component using \citep[see][]{wa98} 
$$A={1\over{[(1-\kappa)^2-\gamma^2]}},\eqno(3)$$
where  $\kappa$ and $\gamma$ for each component  were taken from \citet[][]{fi16}. The obtained amplification is $A$=18.36 (for A), 7.76 (for B), 3.62 (for C), and  1.6 (for D). We measured the fluxes of all four components from observations performed in 2018 because the S/N was best for observations in this epoch. The obtained unlensed luminosities  (taking standard cosmological parameters) for the components at 1350 are $\sim$ 5.1 $10^{44}$ erg s$^{-1}$ (for A),  7.3 $10^{44}$ erg s$^{-1}$ (for B),  7.9 $10^{44}$  erg s$^{-1}$  (for C), and 7.3 $10^{44}$ (for D), which gives an averaged non-lensed quasar luminosity of $\lambda$L(1350)=(6.9$\pm$0.9) 10$^{44}$ erg s$^{-1}$. Using the R-L relation from \citet[][]{ko06}, we obtained that the C IV BLR dimension is  $\sim$ 42 light days, and  using the   
R-L relations for luminous quasars given  by \citet{ka07} and \citet{li18}, we obtained a smaller  C IV BLR dimension of $\sim$ 15-20 light days. The estimated C IV BLR dimension from the C IV line luminosity using the relation given in \citet{ko06} (Eq.(2) given in  the paper) is significantly larger than the one estimated from the continuum. The reverberation relations given in \citet{ka07} and \citet{li18} were derived for high-luminosity quasars, therefore they give some minimum values of the C IV BLR.  The estimated accretion disk dimension in SDSS J1004+4112 is $\sim$9 light days \citep[]{fi16}, which implies that the BLR is at least several times larger than accretion disk. Therefore, the C IV BLR dimension in SDSS J1004+4112 probably is more than some dozen light days.

The microlensing Einstein ring radius (ERR) for J1004+4112 can be calculated as \citep[][]{fi16}
$$ERR=9.1\cdot\sqrt{M\over{0.3M_\odot}}\ \ \textrm{l.\ days,}$$ 
which gives around 9 l. days for a 0.3 solar mass microlens that is assumed to be present in the SDSS J1004+4112 lens system. When we compare the ERR dimensions with the estimated dimensions of the BLR, we cannot expect a significant microlensing magnification of the broad emission lines in this system \citep[see][]{ab02} because the ERR is at least twice smaller than the C IV BLR estimated dimension.  Moreover, the variation in the blue wing of component A seems to occur in a relative short time \citep[see][]{ri04}, therefore microlensing of the whole BLR can be excluded.

As we noted above, the C IV line shapes of non-lensed quasars show a blue asymmetry and/or shift \citep[see, e.g.,][]{ri11,ma19} that indicates an outflow contribution to the C IV broad line fluxes. In principle, different C IV BLR geometries can be considered (spherical, disk-like, outflowing, etc.), but there may be a stratification in the BLR  where one component is disk-like (follows the accretion disk kinematics) and an additional component that originates in an outflow (see Fig. \ref{fig-mod1}).

To explain the amplification in the blue wing of component A, we considered a phenomenological model as shown in the left panel of Fig. \ref{fig-mod1}. The scale in the panel is given in arbitrary units.
The map and scheme of the C IV-emitting region shown in Fig. \ref{fig-mod1} are only an illustration in order to qualitatively explain the CIV blue wing amplification. Because we found that the C IV BLR probably has  larger dimensions (about 40 light days) than the projected  microlens ERR (about 9 light days), the amplification is probably not  due to microlensing of the whole disk-like BLR. The illustration of the disk is shown  in the left panel, which emits an emission line (illustrated as the dashed line in the right panel).
 
There may be a large caustic (shown in the left panel) that slowly crosses the blue part, slightly magnifying the blue wing (the solid line in the right panel). Additionally, the caustic may also microlens the compact part of the emission that comes from the outflowing component (illustrated as a small dashed line in the blue wing;  see the right panel), whichmay vary in a short period. This scenario can provide an explanation of the fast variability in the part of the blue wing only in component A. On the other hand, this can also explain the relatively low intensity of the red wing that is observed in component A with respect to component B.

As we noted above, the HST image of SDSS J1004+4112 shows a point-like object close to image A \citep[][]{sh05} that may be  the source of the spurious emission that contaminates the blue wing of C IV. However, it is hard to explain the nature of an object with  a spectral energy distribution that has a short-wavelength interval that only  contributes to the blue C IV wing intensity. Additionally, we cannot see any contribution to the other lines or additional continuum in the spectrum of image A that shows simultaneous change with the changes in the C IV blue wing. This means that this object probably does not contribute significantly to the blue wing amplification of image A.
 
Finally, we comment on the less outstanding but significant enhancement of the red wing of image D, which may also be related to microlensing. This hypothesis is supported by the presence of microlensing magnification in the continuum of image D.

\begin{figure}[httb]
\centering%
\includegraphics[width=0.45\textwidth]{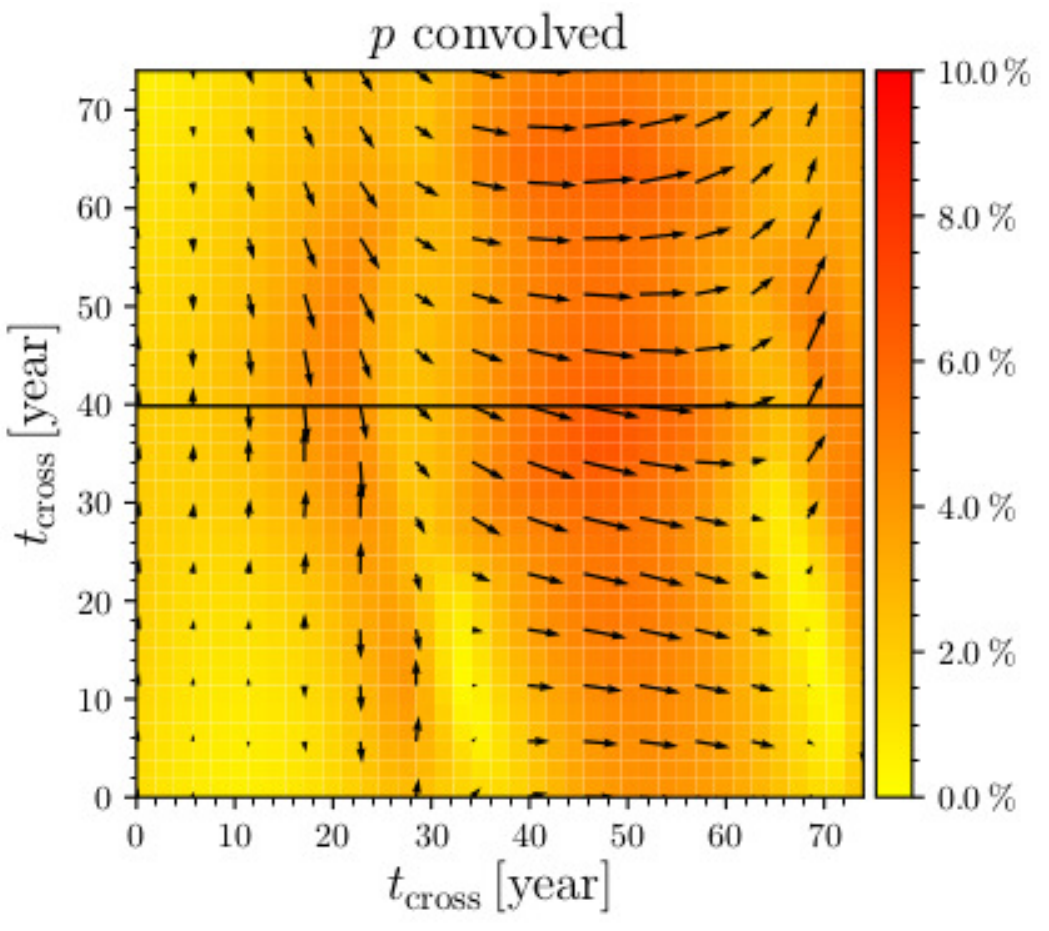}
\caption{ Magnification map (see the appendix) showing the effect of microlensing on the degree of polarization $p$ (given in color) and on the polarization angle $\varphi$ (arrows). The level of polarization is given in percent. Arrows paralell to the Y-axis correspond to $\varphi=0$, and the arrow length corresponds to the $p$ value. The solid black line represents the path of the source, starting from left to right. This map is a subregion of a much larger magnification map shown in the appendix.}
\label{fig:plP_fi}
\end{figure}

\subsection{ Microlensing of the scattering region of the torus}

Assuming that the lensed source has a complex polarization structure, the image polarization can also be expected to be different. Theoretically, several papers have considered the polarization due to  the macro- and microlensing \citep[{see, e.g.,}][]{sc87,bo96}. Observational microlensing effects in the broadband and spectra of the lensed quasar H1413+117 were observed \citep[see][]{ch01,hu15}. In this system, component D showed a higher level of polarization and polarization angle.

The polarization sources in lensed quasars can have a different nature. Based on some observational facts, we expect polar or equatorial scattering to be the main polarization mechanisms in AGNs, with polar scattering dominant in type 2 objects and equatorial scattering in type 1 objects \citep[see, e.g.,][]{sm04,ap15,pop18}. In all images of J1004+4112, broad Ly$\alpha$, C IV, and CIII] lines are observed \citep[see our spectra, as well as those in][]{in03,ri04,gr06,la06,mo12,fi16}. Thus, because SDSS J1004+4112 is a type 1 AGN, we consider equatorial scattering as the dominant polarization mechanism.

\subsubsection{Expected polarization variability during a microlensing event}

As we noted above, variability is observed in the polarization parameters of all components (see Figs. \ref{fig-pol} and \ref{fig-pol1}), but the most prominent variability is observed in component D. The change in polarization correlates with the change in the magnitude of component D, therefore we expect that the microlensing effect can affect the polarization parameters (Stokes parameters, and consequently, the level of polarization and the polarization angle). In order to demonstrate the influence of microlensing on the polarization parameters, we studied the  microlensing effect in the scattering region that is assumed to be located in the inner part of the torus.

To have a realistic model of polarization in AGN, we  modeled the equatorial scattering  using the Monte Carlo radiative  transfer code STOKES  \citep[assuming type 1 AGNs, see more details in][etc.]{go07,ma15,sa18}. For the torus we considered a flared-disk geometry  assuming Thomson scattering in the inner part of the torus. The inner radius of the torus was taken as 0.1 pc ($\sim$13 ERR, see Appendix A), assuming  an optical depth of $\tau=5, $ and the outer radius of the torus scattering region  is 0.2 pc ($\sim$26 ERR). The Stokes parameters were calculated across the entire scattering region 
(for more details, see Appendix A1, see also Fig. \ref{fig:QUI}). We assumed a face-on torus orientation (with respect to an observer), therefore the level of polarization was very low, $\sim$0.2\% of the not microlensed torus.

The polarization rate and angle across the scattering region show a significant gradient. The change in polarization parameters can be detected in the interval of 0.01pc$\approx$ 12 l.d, which is comparable with an ERR of a star with 0.3 M$\odot$ ($\sim 9$ l.d.).

Additionally, we generated a microlensing map for image D, taking the estimates for convergence and shear given in \citet{fi16}. The dimensions of the map are 4000$\times$4000 l.d.$^2$, with a resolution of 1 light day (1pix=1 l.d.), which is equivalent to 241.6$\times$241.6 ERR$^2$.

This map was convoluted with the Q and U Stokes parameter maps, which have dimensions 476$\times$476 l.d, with the same resolution 1pix=1 l.d. After convolving the Stokes parameters with the microlensing map, we calculated p and $\varphi$ for each pixel according to Equation A2. Fig. \ref{fig:plP_fi} is a zoom-in of the larger polarization map obtained after convolution. A detailed description of our simulations is given in Appendix A.

To gain an impression of the timescale, we converted the polarization level and angle maps into the standard timescale using \citep[see][]{tr04,jov08}
$$ t_E \approx (1+z_l){\mathrm{ERR}\over {v_{\perp}}},$$
where $\mathrm{ERR}$ is the point-like ERR projected onto the source plane, $v_\perp$ is the relative transverse velocity  that is typically $v_\perp\sim$ 600 km s$^{-1}$ \citep[see][]{tr04}, and $z_l$ is the cosmological redshift of the lens.

\begin{figure}[httb]
\centering%
\includegraphics[width=0.45\textwidth]{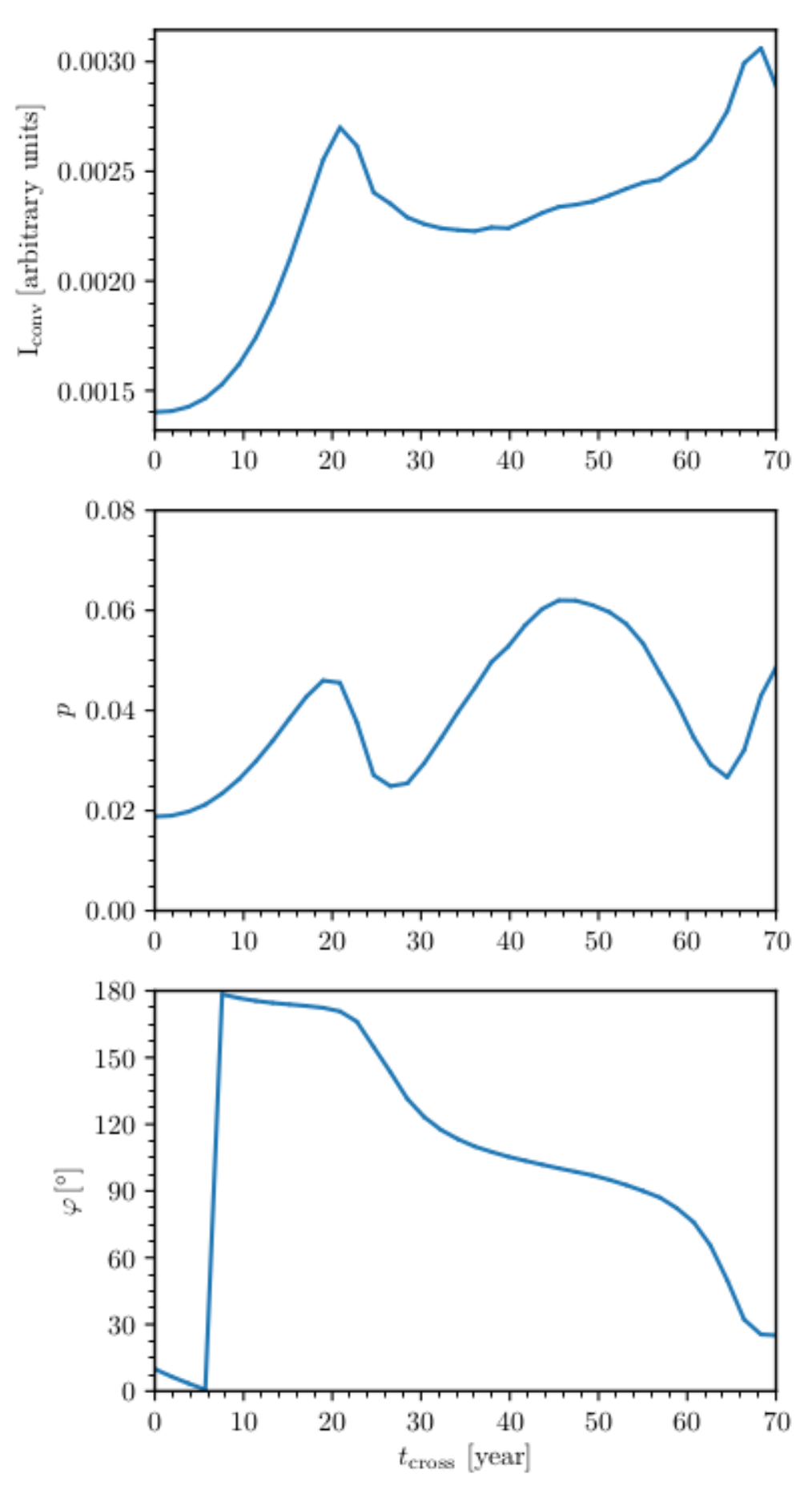}
\caption{Modeled changes in intensity (up), degree of polarization $p$ (middle), and polarization angle $\varphi$ ( bottom) corresponding to the crossing path shown as a solid line  in Fig.\,\ref{fig:plP_fi}.}
\label{fig:plP_fi_cross}
\end{figure}

In Figure \ref{fig:plP_fi} we present the magnification map corresponding to the polarization rate and polarization angle. The solid line in Fig. \ref{fig:plP_fi} represents the source path across the map, which results in the change of  parameters that is shown in  Fig. \ref{fig:plP_fi_cross}. The first plot in Fig. \ref{fig:plP_fi_cross} presents the change in intensity due to microlensing. The total intensity accounts for the radiation of the point source plus the radiation scattered off the torus. The second plot shows the change in the polarization (that can be from 2\% to 6\%), and the third plot shows the change in polarization angle. Fig.  \ref{fig:plP_fi_cross} shows that the observed polarization  of component D can be qualitatively described by the model, and the polarization level and the angle of polarization can change significantly during a microlensing event. Additionally, in some cases, microlensing can strongly increase the polarization rate (by a factor 10, see Fig. \ref{fig:plP_fi_cross}). However, using our model and calculating the probably density function for the polarization rate (see in Appendix Fig. A5), we found that our model (microlens and source) gives that the most probable microlens rate is about 1\%, and has a relatively high probability until 4\%. This agrees with the observed polarization variability in component D.

\begin{figure}[httb]
\centering
\includegraphics[width=0.45\textwidth]{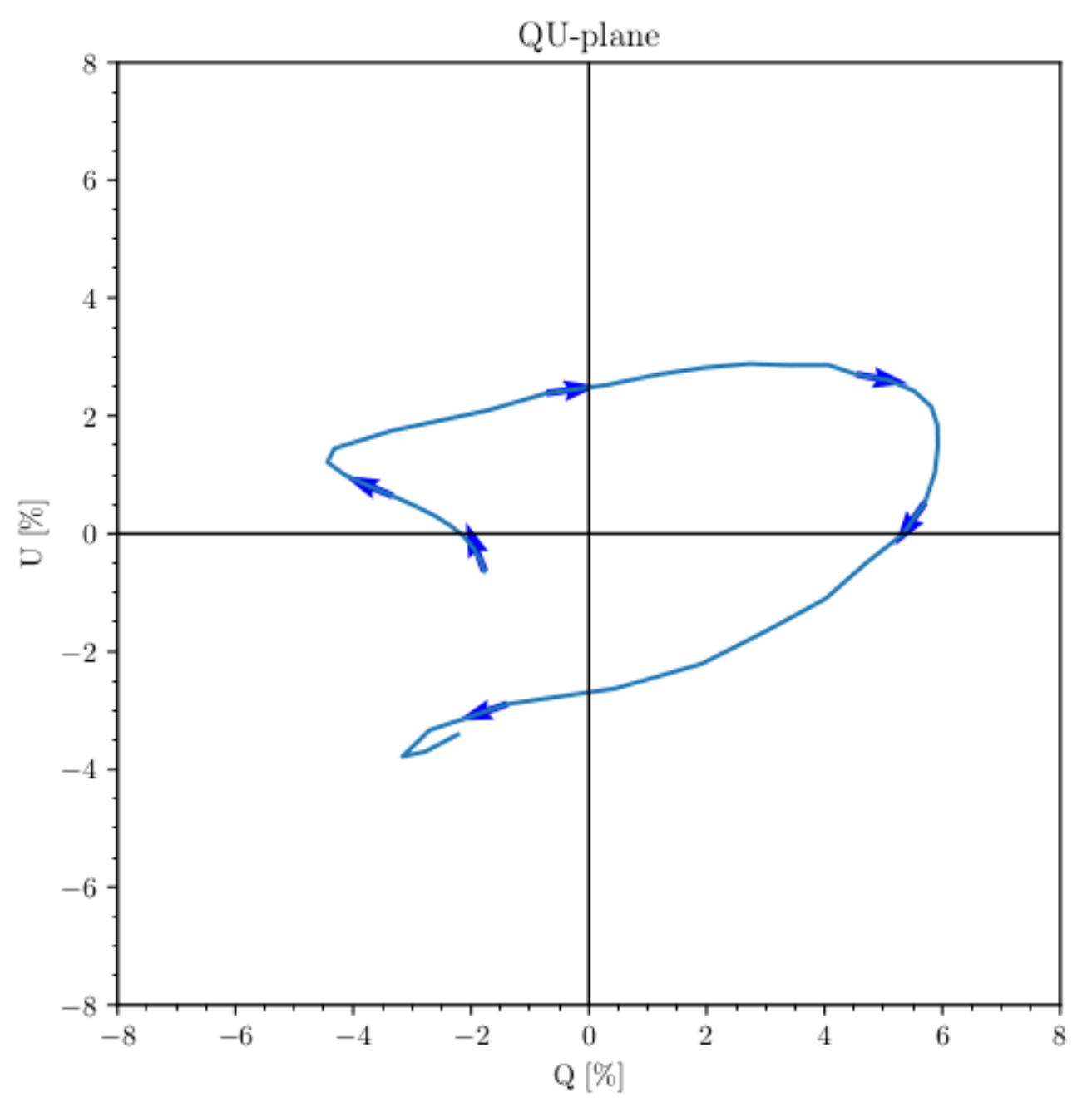}
\caption{Variation in Stokes parameters Q and U along the crossing path defined in Fig.\,\ref{fig:plP_fi}.
Arrows denote the direction in which Q and U evolve as the source crosses the path on the microlening map, starting from left to right.}
\label{fig:QU_plane}
\end{figure}

\
In Fig. \ref{fig:QU_plane} we present the variability in the UQ plane. The changes in the UQ plane can be significant during a microlensing event. Consequently, the polarization angle can change from 0 to 180 degrees and the level of polarization can reach 10\%. In our simulation  we did not see any significant correlation between the change in polarization angle and the polarization level. The amplification in intensity coincides with the change in polarization parameters, but the correlation  between the behavior of these changes is not significant (i.e., the maximum intensity does not correspond with the maximum polarization level, etc.).

The difference between  polarization angle of image C is very different from the polarization angle observed in images A and B. This difference  may be also due to microlensing, but at the beginning of microlensing,   without any sign of a strong change in polarization degree. Fig. \ref{fig:plP_fi_cross} shows that the changes in polarization angle can be strong (third panel of Fig.  \ref{fig:plP_fi_cross}) without strong changes in polarization rate (middle panel of Fig. \ref{fig:plP_fi_cross}). However, this needs to be confirmed by future observations in polarized light of this component.

\subsubsection{Effect of macrolensing on image polarization}

We also explored the possibility that the different locations in the QU plane of the averaged polarization values for the different images were caused by macrolensing through a different transformation of the source. To do this study we fit a simple SIS$+\gamma$e model to the positions of the four images of J1004+4112. As far as images A and B are relatively close (in fact in near infrared observations taken with the HST it can be seen that for enough large sources the images merge into one arc) we can think that the central position of the source is not far away from a macro-caustic. We apply this model to the 2D distributions of Q and U Stokes parameters of the torus described in section 4.2.1. to compute their lensed images and to calculatethe histogram of polarizations for the  four images. In the case of the detached C and D images, we find no differences with respect to the source histogram. That is, in images C and D, the polarization changes cannot be attributed to macrolensing. This result seems easy to explain. Lensing acts like a linear transformation in a region in space if its size is small enough. The source is very small (0.05 mas, according to the torus dimensions described in section 4.2.1.) and each surface element is transformed under the same linear transformation, therefore the source histogram does not change under macrolensing. In the case of images A and B, which appear merged in the near-infrared, the situation may be different. If the source is large enough to be crossed by a caustic (a possibility, although the torus is so small that it seems unlikely, but not impossible because the location of the source with respect to the caustic depends on the lens models), lensing will produce two images of the whole source and two additional images of only the inner part of the caustic part of the source. Under these circumstances, macrolensing could change polarization. However, A and B show no significant differences in the QU plane. Thus, there is no reasons to assume that the caustic intersects the torus, confirming that it is far smaller than the source that causes the near-infrared arcs. Microlensing consequently is a more plausible explanation for the relative shifts between images in the QU plane.

\section{Conclusions}

We presented spectroscopic and polarimetric observations of the lensed quasar J1004+4112 obtained with the 6m telescope of SAO RAS. We analyzed the observed spectra, and the C IV blue line bump in component A may be explained with a phenomenological model for the BLR that includes an outflowing component. Additionally, we studied the effect of  macro- and microlensing on the polarized light caused by Thomson scattering in the inner part of the torus. Based on our analysis, we can outline the following conclusions:

\begin{itemize}
 \item The CIV blue bump seen only in image A that has been reported in earlier observations  is also present in the three epochs (2007, 2008, and 2018) of our observations. The CIII] line profile of image A is similar to the line profiles observed in the other three images. To explain this effect, we propose that the outflow contributes to the blue wing. When we assume that the C IV BLR is about several dozen light days, the outflowing region should be more compact (several light days) in order to be microlensed by a microlens ERR of $\sim $ 9 light days, as estimated for the SDSS J1004+4112 lens system.  
 \item We observed the polarized light of J1004+4112. We find that the averaged positions on the UQ-plane of the different images are different. Components A and B have an average  polarization angle of about 40-50 degrees, while the C and D components have an averaged polarization angle of about 140 degrees. The relatively small size of the torus makes an explanation of these differences related to macrolensing unlikely, which basically acts like a linear transformation of the source.
 \item Significant variability of the polarization is observed only in component D. Simulation of the microlensing of a scattering region in the inner part of the torus can qualitatively explain the observed changes in the UQ-plane, as well as the change in polarization level and the polarization angle of image D observed in the 2014-2017 period.
\end{itemize}

Additionally, we presented a qualitative model of microlensing of a disk-like plus emitting outflow BLR to explain the observed magnification of the CIV blue wing of component A. This needs to be confirmed in future spectroscopic observations.

\section{Acknowledgments}

This work is supported by the Ministry of Education, Science and Technological Development of R. Serbia (the project 146001: ``Astrophysical Spectroscopy of Extragalactic Objects'') and Russian Foundation for Basic Research  (grant No. 19-32500). The observations at the SAO RAS 6 m telescope were carried out with the financial support  of the Ministry of Science and Higher Education of the Russian Federation. Dj. Savi\'c thanks the RFBR for the realization of the three months short term scientific visit at SAO funded by the grant 19-32-50009. We would like to thank to the referee for very useful comments.
%
%

%



\begin{appendix} 

\section{Microlensing of an equatorial scattering region in AGN}

Characteristic polarization in type 1 AGNs is caused by equatorial scattering in the innermost part of the torus \citep[see][]{pop18}. Therefore, here we consider microlensing of the equatorial scattering region, which is assumed to be located in the inner part of the torus.

\subsection{Modeling polarization: equatorial scattering in the torus}
To simulate the equatorial scattering in the inner part of torus, we applied the 3D Monte Carlo radiative transfer code \textsc{stokes} of \cite{go07}. The code was initially built for studying optical and UV polarization induced by electron or dust scattering in type 1 AGNs. The code follows the trajectory of each photon from its creation inside a user-defined emitting region, until it finally reaches the distant observer. During this time, the photon can undergo a chain of scattering events with its polarization state changed and recorded after each scattering. If there is no scattering region in the photon path, the photon polarization state is recorded by a web of virtual detectors that surround the system. The program ends after very many photons (typically $10^{11}$ or more) are registered and the obtained statistics is good. We used the \textsc{stokes} version 1.2, which is publicly available\footnote{http://www.stokes-program.info/}. The advantage of this version is that it allows us to run the code in imaging mode, thus producing images projected onto the observer's plane of the sky. We adopted the same convention as \cite{go07} for the polarization angle: $\varphi$ parallel to the $y-$axis has a value $\varphi = 90^\circ$.

\begin{figure*}[httb]
\centering
\includegraphics[width=0.7\textwidth]{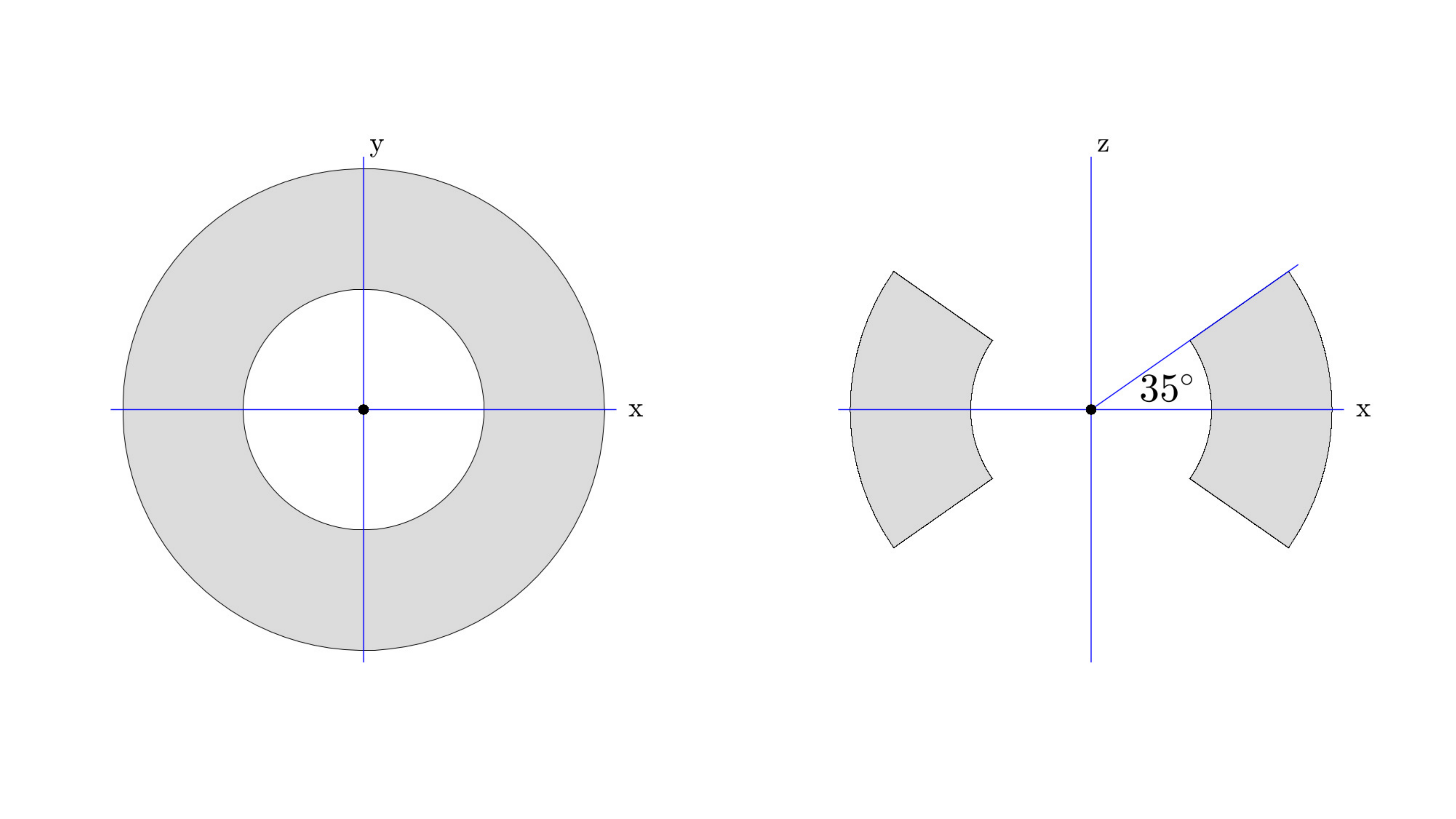}
\caption{Scheme of the torus geometry. Face-on (left) and edge-on (right). The continuum source is considered to be a point-like source in the center of the torus.}
\label{fig:map1}
\end{figure*}

\begin{figure*}[httb]
\centering
\includegraphics[width=.90\textwidth]{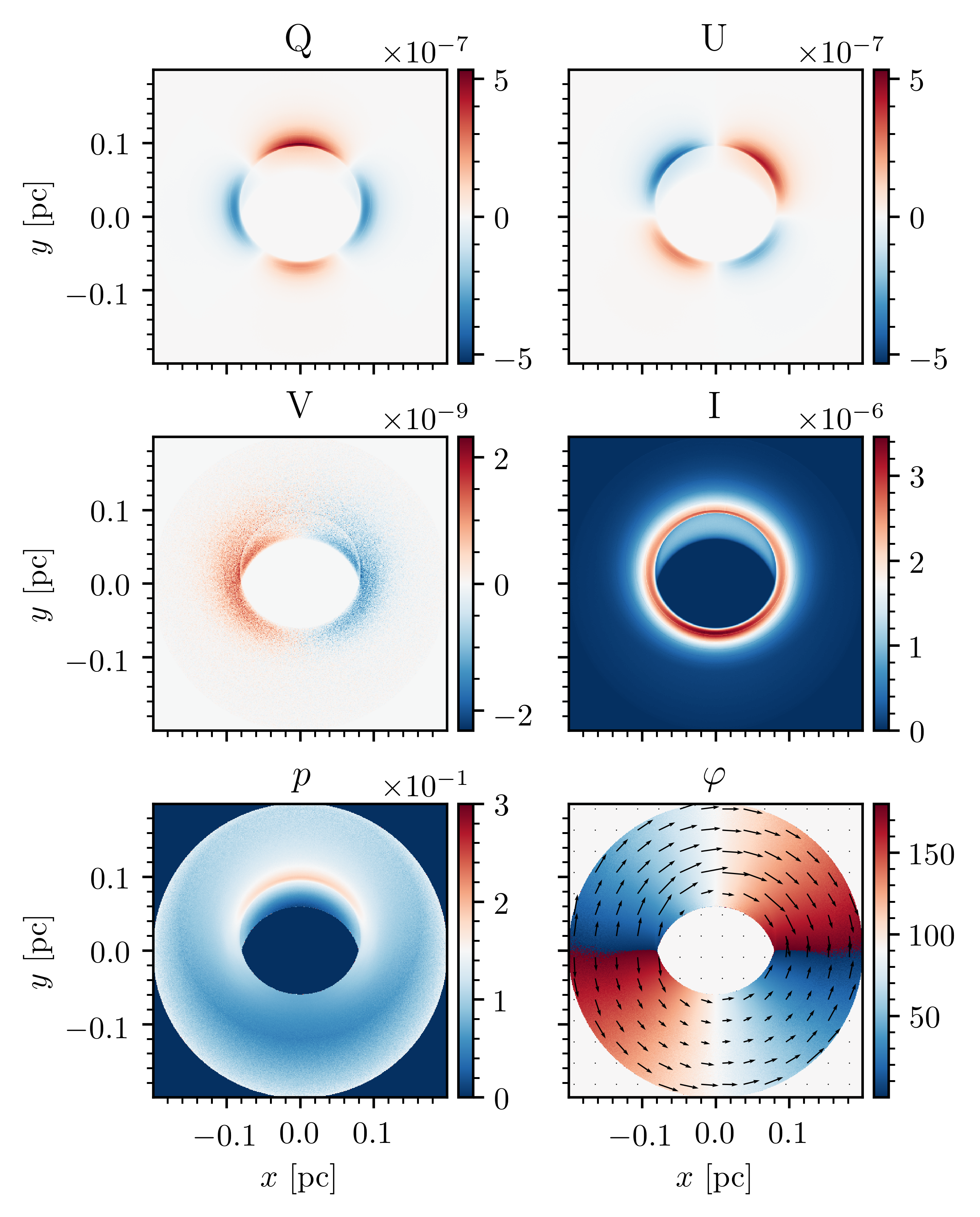}
\caption{Stokes parameters Q (top left), U (top right), V (middle left), and I (middle right);  degree of polarization $p$ (bottom left), and polarization angle $\varphi$ (bottom right). Stokes parameters are normalized with respect to the total value of I integrated over the entire torus and including the contribution from the central source. The degree of polarization is given in fractions. The polarization angle is computed with respect to the $y-$axis. For better visualization, $\varphi$ is also shown as a vector with sizes corresponding to $p$.}
\label{fig:QUI}
\end{figure*}

We modeled the continuum polarization, which due to Thomson scattering  does not depend on wavelength. A simple AGN geometry was modeled, considering the accretion disk radiation in the continuum as an isotropic point-like source of radiation. The spectral energy distribution (SED) is given by a power law $F_c \propto \nu^{-\alpha}$, where $\alpha$ is the spectral index. The  dusty torus was modeled using a flared-disk geometry with a half-opening angle of $35^\circ$ when measured from the equatorial plane. Considering a high radial optical depth and a torus size of about a few parsec, it is sufficient to treat equatorial scattering only at scales that are a few times larger than the mean free path of the photon $\overline{l}$. For a homogeneous dust distribution, it can be shown that the photon mean free path only depends on the size of the torus and the total optical depth as

\begin{equation}
\overline{l} = \frac{L}{\tau},
\end{equation}
where $L$ is the difference between the inner and outer edge of the torus, and $\tau$ is the total optical depth. Photons that reaching farther in the torus interior have a very high probability of being absorbed by dust particles.
  
The inner torus radius can be estimated by reverberation mapping measurements \citep[see, e.g.,][]{su06,ki11,ko14}. As e.g. \citet{ko14} using the V absolute luminosity,  which in our case yields a torus inner radius of ~0.1 pc. Knowing that the dusty torus usually spans a parsec scale \citep[e.g., 3 pc for Circinus galaxy, see][]{st19} and that the total radial optical depth in the equatorial plane for the entire torus size in V band  is high $\sim$150 \citep[see][]{ro18}, we obtain that one mean free path of the photon for scattering is on the torus inner radius of $\sim$ 0.02 pc. The probability of the photons of being absorbed after five lengths of mean free path is high, and we can assumed that the greatest part of photons are scattered within  $5\times0.02$ pc from the inner wall of the torus, while the rest of the photons that penetrate deeper are absorbed. Taking this into account, we set the torus inner and outer boundaries to 0.1 pc and 0.2 pc, respectively. For this segment, we adopted an optical depth of $\tau = 5$ in such way that the dust concentration radially decreased as $n_{\mathrm{dust}} \propto r^{-1}$  \citep[][]{sm04}. An illustration of the model is shown in Fig.\,\ref{fig:map1}. We used the Milky Way dust prescription by \citet{ma77}, which is implemented in \textsc{stokes} by default.

The images corresponding to the modeled Stokes parameters of the torus are shown in Fig. \ref{fig:QUI}. From top to bottom, we show images of the spatial distributions of the Stokes parameters I, Q, U, and V, as well as the degree of polarization $p$ and the polarization position angle $\varphi$.

The system is viewed from a nearly pole-on viewing inclination $\theta = 15^\circ$. The Stokes parameter Q is shown in Fig.\,\ref{fig:QUI} (top left panel). It shows axis-symmetry with respect to the $y$-axis, and the total Q value integrated over the torus is greater than zero because the far inner side contributes more than the near inner side of the torus. The light scattered of the far inner side of the torus is seen more in 'reflection', therefore it contributes more to the linear polarization than the near inner side, which is seen more in 'transmission'.

The Stokes parameter U (Fig.\,\ref{fig:QUI}, top right panel) shows the same behavior as Q, with the  difference that it is antisymmetric with respect to the $y$-axis. Therefore the net U parameter integrated over the whole image should be zero because it  is within the Monte Carlo uncertainty of our simulations. Dust scattering can produce a low degree of circular polarization. For academic purposes, it is therefore worth mentioning the Stokes parameter V (Fig.\,\ref{fig:QUI}, middle left panel). It is antisymmetric with respect to the $y$-axis, with the left part taking positive values and the right part taking negative values. As in the case of U, the expected total value is zero. Because the values of V are two orders of magnitude lower than the values of Q and U, we focused on linear polarization  and neglected circular polarization. The parameter I is shown in Fig.\,\ref{fig:QUI} (middle right panel). As expected, the unpolarized light is mostly seen in transmission from the upper closer side of the torus. We point out that the unpolarized light coming from the central source directly to the observer (omitted in the plot) contributes roughly 90\% to the total unpolarized flux. The degree of linear polarization is shown in Fig.\,\ref{fig:QUI} (bottom left panel). The polarized radiation predominantly comes from the scattering of the far upper inner side of the torus (bluish crescent shape). This is not to be confused with the unusually high $p$ values because they were calculated for each pixel. When the central source is taken into account, the net $p$ is lower than 1\%, as expected for a nearly pole-on view (or type 1 AGNs). The polarization angle is shown in Fig.\,\ref{fig:QUI} (bottom right panel). The angle follows the shape of the torus. Along the constant line $y = 0\,\mathrm{pc}$, the discontinuity between the upper and the lower side of the torus is visible. The upper side shows increasing values from $0^\circ$ to $180^\circ$, while the lower side behaves in the opposite way: decreasing values from $180^\circ$ to $0^\circ$. Because the net U value is 0, the net $\varphi$ integrated over the whole torus is $90^\circ$.


\subsection{Microlensing model}

To determine the  microlensing parameters the J1004+4112 image positions were fit with the model of singular isothermal sphere plus external shear \citep[SIS+$\gamma$e, see][]{fi16}. The model gives convergence $\kappa=0.71$ and shear $\gamma=0.83$ for image D. We assumed a surface mass density in stars of 10\% according to \citet[][]{med09}.

\begin{figure}[httb]
\centering
\includegraphics[width=0.5\textwidth]{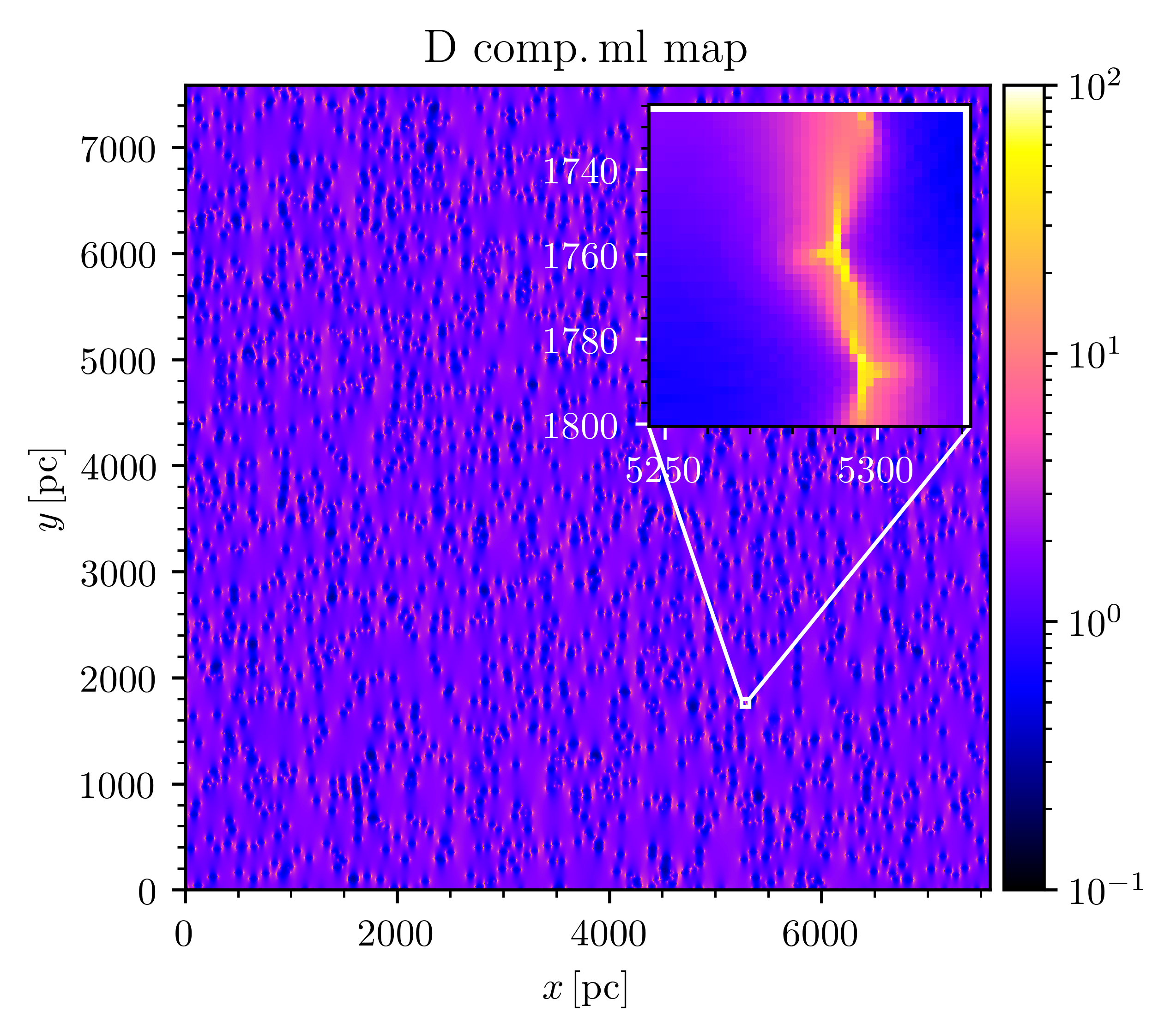}
\caption{ Microlensing magnification map of image D of the gravitational lens system SDSS J1004+4112.}
\label{fig:map}
\end{figure}

In order to estimate the influence of gravitational microlensing on the scattering region, we computed microlensing magnification maps using the inverse polygon mapping method described in Mediavilla et al. (2006) and \cite{med11}. In Fig.\,\ref{fig:map} we plot the map for image D (4000$\times$4000 pixels),  calculated using the following parameters: convergence, $\kappa=$0.71,  shear,  and $\gamma=$0.83, and with a microlens masses of 0.3$M_\odot$. The resolution of the maps is one light day. For the lens and source redshifts, we took $z_d=0.68$ and $z_s=1.734$.  We adopted standard cosmological parameters ($H_0=71$, $\Omega_m=0.27$ and $\Omega_{\Lambda}=0.73$.

\subsection{Microlensing magnification of polarization parameters}

\begin{figure*}[]
\centering
\includegraphics[width=0.45\textwidth]{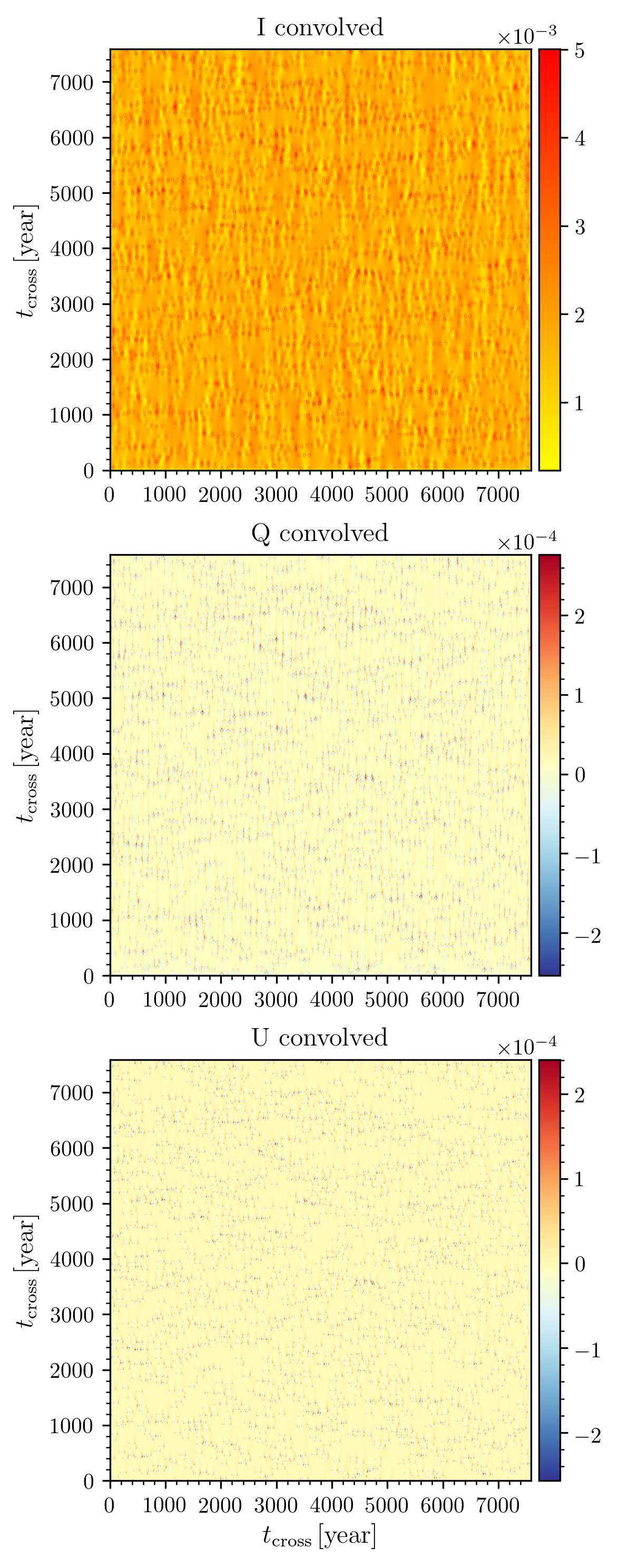}
\includegraphics[width=0.45\textwidth]{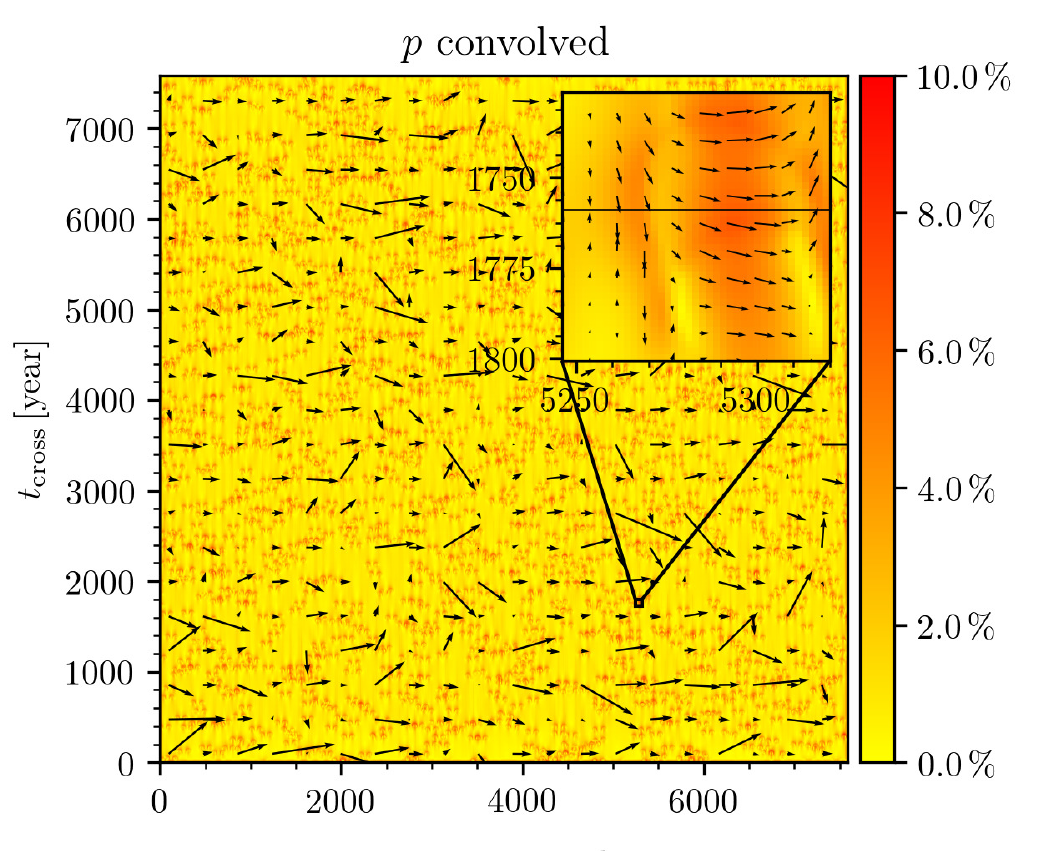}
\caption{Left: Convolution of the total intensity (top panel) and  Stokes parameters Q (middle panel) and U (bottom panel) with the modeled magnification map for component D. Right: 2D distribution of the degree of polarization, $p$ (coded in color levels and in the length of the arrows) and polarization angle $\varphi$ represented using arrows (arrows parallel to the Y-axis correspond to zero angle). The inset at the top is the map we used to calculate the polarization amplification (see text and Fig. 9).}
\label{fig:plQUIconv}
\end{figure*}

A magnification map can be combined with the images of the source that correspond to the polarization parameters ($Q$, $U,$ and $I$ - see Fig.\,\ref{fig:QUI}) to evaluate the influence of microlensing. We point out again here that parameter $I$ consists of two components: the dominant central source, and the fainter scattering region. In order to do this, we convolved the magnification map of image D with the modeled images of the polarization parameters of the source (see Fig. \ref{fig:QUI}), obtaining three separate convolutions that we present in Fig.\,\ref{fig:plQUIconv} (left three panels). Here blue and dark blue present areas with magnified polarization parameters, while red and dark red places designate negative amplification or their absence. Convolution computes the influence of gravitational microlens on the source (in our case, the generated Stokes parameters) at any position on the magnification map. In order to compute the variations in polarized light intensity when the microlens passes over the source, we therefore only need to extract the map slice corresponding to the trajectory. In this way, we computed light curves that represent the effect of microlensing on the polarization parameters Q, U, and I.

\begin{figure*}[]
\centering
\includegraphics[width=0.45\textwidth]{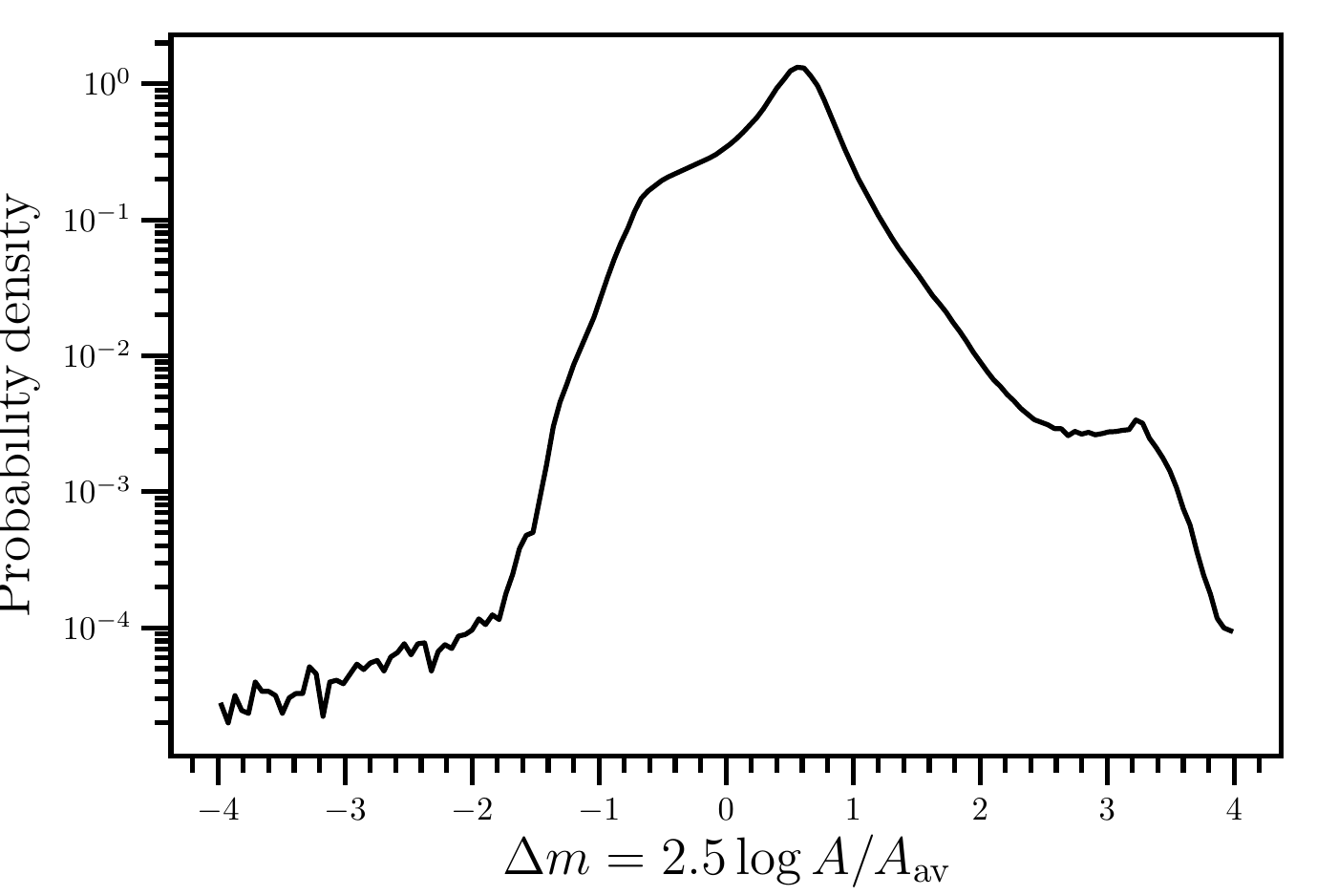}
\includegraphics[width=0.45\textwidth]{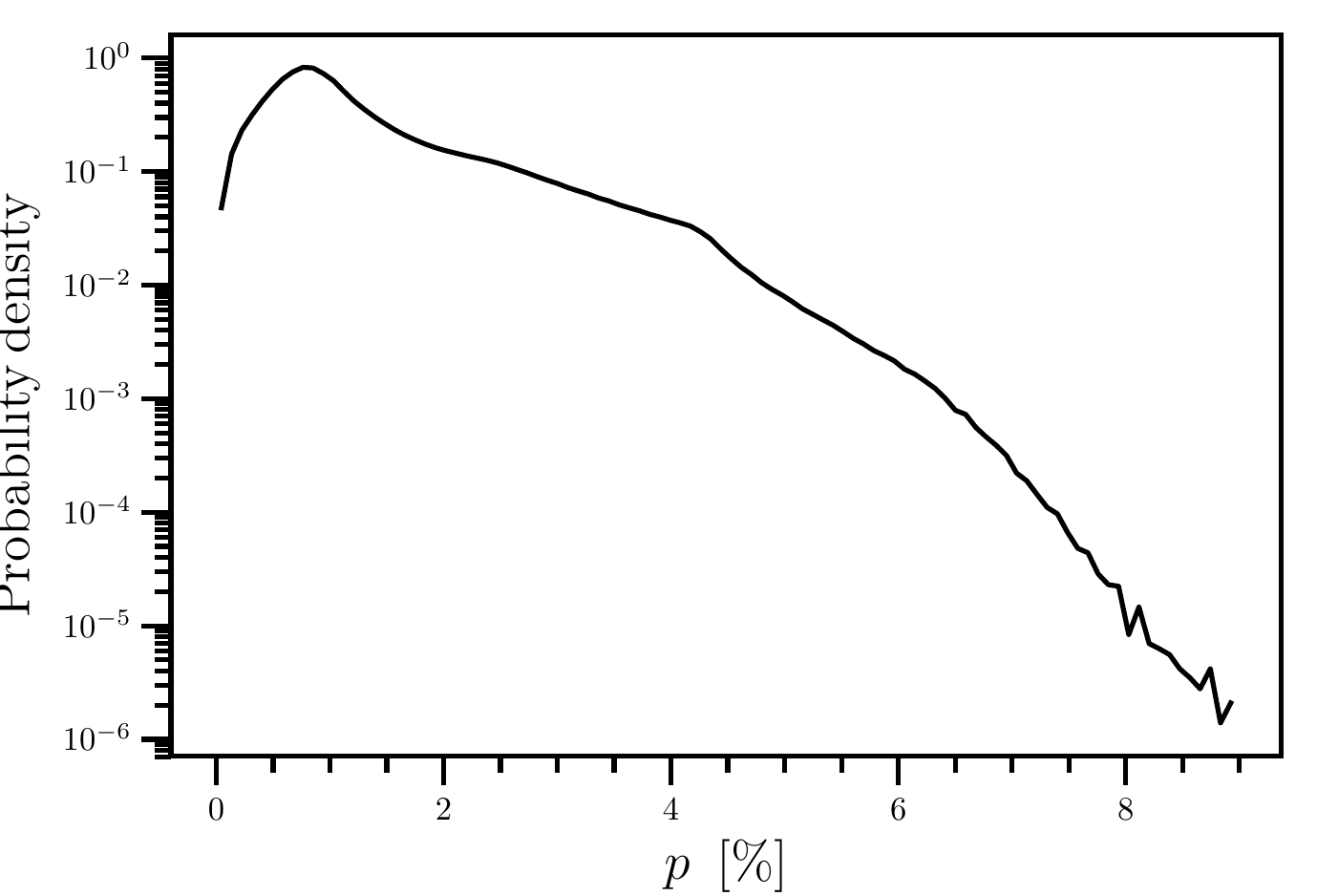}
\caption{Probably density functions for the amplification (left) and polarization rate (right).}
\label{fig:pdf}
\end{figure*}

In order to compute the polarization $P$ and the angle of polarization $\varphi$ , we used the following equations:
\begin{equation}
P=\sqrt{Q_r^2+U_r^2}  \qquad \varphi=\frac{1}{2}\arctan\frac{U_r}{Q_r},
\label{eq:P_varphi}
\end{equation}
where all three parameters (Q, U, and I) are obtained from  the convolved maps, while parameters Q and U are additionally scaled to the value of parameter I (pixel by pixel), therefore they have the index $r$.

In Fig. \ref{fig:plQUIconv} (right panel) we show the map of $p$ (color corresponds to intensity) and $\varphi$ (represented by arrows, where the intensity of arrows represents the degree of polarization). The figure shows that to convolve the torus, which has large dimensions, with the magnification map, we used the entire map, and to compare changes in polarization parameters in a period of several years, we took a small part (zoomed in Fig. \ref{fig:plQUIconv}) on which we simulated the transit of a source (solid line in Fig. \ref{fig:plP_fi}).

Additionally, we calculated the probability density functions \citep[PDFs, see][]{w92} for the amplification and degree of polarization (see Fig. \ref{fig:pdf}) of component D. In Fig. \ref{fig:pdf} we present the PDF for the amplification as a function of magnitude,
$$
 m = 2.5\log_{10}(A/A_{\mathrm{av}}),
$$
where $A$ is the total magnification and $A_{\mathrm{av}}$ is the average magnification given by Eq. (3) The peak around $\Delta m\approx 1$ in the magnification is clear, and the polarization rate has a maximum around 1\%, but there is also a reasonable  probability for a polarization rate between 1\% and  4\%.
\end{appendix}

\end{document}